\documentclass[12pt]{article}
\usepackage{amsmath,amssymb}
\usepackage{comment}
\usepackage{bm}
\usepackage{color}
\usepackage{graphicx}
\usepackage[dvipdfmx]{epsfig}
\usepackage{epsfig}
\usepackage{braket}
\usepackage{color}
\usepackage{hyperref}
\newcommand{\tmop}[1]{\ensuremath{\operatorname{#1}}}
\newcommand{\nobracket}{}

\newcommand{\bea}{\begin{eqnarray}}
\newcommand{\eea}{\end{eqnarray}}

\newcommand{\nn}{\nonumber}

\begin{document}
\title{
\begin{flushright}
\ \\*[-80pt] 
\begin{minipage}{0.2\linewidth}
\normalsize
HUPD-2105 \\*[50pt]
\end{minipage}
\end{flushright}
{\Large \bf 
A model with light and heavy scalars \\ in view of the effective theory
\\*[20pt]}}

\author{ 
\centerline{
Apriadi Salim Adam$^{1}$\footnote{E-mail address: apri014@lipi.go.id},
Yuta Kawamura$^{2}$\footnote{E-mail address: yuta-kawamura@hiroshima-u.ac.jp}, and
 Takuya~Morozumi$^{3,4}$\footnote{E-mail address: morozumi@hiroshima-u.ac.jp}
}
\\*[20pt]
\centerline{
\begin{minipage}{\linewidth}
\begin{center}
$^1${\it \normalsize
Research Center for Physics, National Research and Innovation Agency (BRIN),\\
Serpong PUSPIPTEK Area, Tangerang  Selatan 15314, Indonesia} \\*[5pt]
$^2${\it \normalsize
Graduate~School~of~Science,~Hiroshima~University, \\
Higashi-Hiroshima~739-8526,~Japan} \\*[5pt]
$^3${\it \normalsize
Physics Program, Graduate School of Advanced Science and Engineering,~Hiroshima~University, \\
Higashi-Hiroshima~739-8526,~Japan} \\*[5pt]
$^4${\it \normalsize 
Core~of~Research~for~the~Energetic~Universe,~Hiroshima~University, \\
Higashi-Hiroshima~739-8526,~Japan} \\*[5pt]
\end{center}
\end{minipage}}
\\*[50pt]}

\date{
\centerline{\small \bf Abstract}
\begin{minipage}{0.9\linewidth}
\medskip 
\medskip 
\small
The low energy effective potential for the model with a light scalar and a heavy scalar is derived. 
We perform the path integration for both  heavy and light scalars and  derive the low energy 
effective potential in terms of only the light scalar. The effective potential is  independent of the renormalization
scale approximately.
By setting  the renormalization scale equal to the mass of the heavy scalar, one finds the corrections
with the logarithm of the ratio of the two scalar masses.  
The large logarithm is summed with the renormalization group (RG) and the RG 
improved effective potential is derived.
The improved effective potential includes the one-loop correction of the heavy scalar 
and the leading logarithmic corrections due to the light scalar.
We study the correction to the vacuum expectation value of the light scalar and the dependence on the mass of the heavy scalar.
\end{minipage}  
}
\begin{titlepage}
\maketitle 
\thispagestyle{empty}
\end{titlepage}
\section{Introduction}
\label{sec1}
Among various models beyond the standard model, the models including particles with large mass hierarchies 
lead to two types of the corrections to the low energy effective potential which is 
obtained by integrating the heavy particles. 
One is large logarithmic corrections to the dimensionless coupling constants
and they have the form of the logarithm of the
large energy scale ratios. 
Another type of the corrections occurs for the mass parameters and the cosmological constants.
The correction to the light scalar mass  often has the form proportional to
the heavy scalar mass. This  type
of correction  destabilizes the initial mass scale set in the tree level approximation.
The correction to the cosmological constants is also sensitive to
the masses included in the model. 


In this paper,  we study the model with two scalars and derive the
low energy effective potential which contains afore-mentioned corrections. 
The model which we consider is a toy model which explains the origin of the small Yukawa coupling to neutrinos
\footnote{In the framework of the standard model gauge group, the two Higgs doublet model with the spirit similar to the present model  has been studied in \cite{Davidson:2009ha, Gabriel:2006ns}.
The constraints from electroweak precision data to these models are discussed in \cite{Machado:2015sha}.}.
In the present model, the light scalar which corresponds to the standard model-like Higgs does not directly couple to the neutrino.
The  light and heavy scalars have a mass mixing term and 
the neutrino Yukawa coupling to the light scalar is generated after the heavy scalar  is integrated out. 
Then the Yukawa coupling is naturally suppressed by a factor; the small mass mixing term between heavy and light scalars divided by heavy scalar mass.

To suppress Yukawa coupling of the neutrinos, we need to increase the heavy scalar mass provided that the small mass mixing term is fixed.  
By increasing the heavy scalar mass, the couplings, the masses and the cosmological constants in the low energy are changed.
To quantify the change, we derive the low energy effective potential for the model. 
With the effective theory approach, one can show how the parameters at the low energy are related to
those of the full theory\cite{Manohar:2020nzp}.  
The vacuum expectation value (vev) of the lighter scalar is also sensitive to the full theory parameters
through the low energy parameters of the effective theory.  We study the dependence of the vev on the heavy scalar mass. If one requires the 
correction to the vev lies within some small range,  one can set the upper bound on the heavy scalar mass.  



To derive the effective potential, we  introduce the source for the light scalar only.  
One first integrates both light scalar and heavy scalar under the constant expectation value of the light scalar.
With this method, one particle irreducibility of the light scalar is maintained while the heavy scalar exchanged diagram 
that is no longer one-particle irreducible  is also included.   
 Then we employ the approximation of the effective potential
by keeping the first few series of the small expansion parameters. 
 Then the obtained  effective potential is also independent of the renormalization scale approximately. 
 
 When there is a large mass difference between the light scalar and the heavy scalar, the
 logarithm of their mass ratio is large and they should be resummed. 
 For this purpose, we derive the  low energy effective theory
for the light scalar and apply the renormalization group (RG) equation in the leading logarithmic approximation. 
Then we derive the RG improved effective potential.

The rest of this paper is organized as follows. In section 2, we present the action for the  model with heavy and light scalars. The definition of the effective action
for the light scalar is given in section 3.  In section 4, we integrate out the light and the heavy scalars. In section 5, the counter terms and the renormalized 
effective potential are derived.  In Section 6, the RG improved effective potential is obtained and Section 7 is devoted to summary and discussion. 
\section{The action for the model in terms of the renormalized quantities and fields}
\label{sec2}

We present a simple toy model which leads to a tiny Yukawa coupling to neutrinos with the light scalar. 
 The light scalar does not couple to the neutrino directly and it couples through the second scalar with the mixing term.
The mass of the second scalar is large and it directly couples to  the neutrino. We denote
the light scalar as $\rho_1$ and the heavy scalar as $\rho_2$ respectively . The neutrino is denoted as $n$. 
The action is written in terms of bare fields $\rho_{0i}$ , bare masses $m_{0i}$ and bare couplings $\lambda_{0i}$ as,
\bea
S =& & \int d^d x \left(- \frac{1}{2}  \sum_{i=1}^2 \rho_{0i}\left( \Box + m_{0i}^2
	+ \frac{\lambda_{0i}}{2} \rho_{0i}^2 \right) \rho_{0i} - \frac{\lambda_{03}}{4}  \rho_{01}^2 \rho_{02}^2 
	-  (y_0 \bar{n_0} n_0 + m_{012}^2 \rho_{01}) \rho_{02} \right.  \nn \\
&& \left. + \sum_{i=1}^{2} h_{0 i} m^4_{0i} +h_{0 12} {m_0}_{12}^4 + 2 h_{03} m^2_{01} m^2_{02}  \right), \label{Action_bare}
\eea
where $m_{012}^2$ is the bare mixing mass.   $y_0$ is the Yukawa coupling between the neutrino and the heavy scalar.
The last lines of Eq.(\ref{Action_bare}) correspond to the cosmological constants which  consist of the mass parameters of the model \cite{Bando:1992wy}. 
We do not include the kinetic term for the neutrino and its quantum correction is not considered.  

The model is renormalizable by imposing the following two $Z_2$
symmetries.  One of them is an exact symmetry called as $Z_2$ and another is a softly broken symmetry called as $Z^\prime_2$. The charge assignment is shown  in 
Table \ref{tab:tbA}.
Two scalar fields $\rho_1, \rho_2$ and right-handed neutrino $n_R$ have odd $Z_2$ parity and the left-handed neutrino $n_L$ has even $Z_2$ parity. 
Under the softly broken $Z^\prime_{2}$, only the scalar field $\rho_1$ has odd parity. 
With this assignment, only the heavy scalar $\rho_2$ has the Dirac type Yukawa coupling to the neutrino.  
In the scalar potential, the cubic interactions of scalars are forbidden by $Z_{2}$
symmetry.  About quartic couplings of the scalar field, one must have even number of each scalar field due to the $Z^\prime_2$ symmetry.   About the quadratic part, the mixing term which is proportional to $\rho_1 \rho_2 $ is allowed, and it softly breaks $Z^\prime_2$ symmetry. 
Concerning the Yukawa interaction between neutrinos and scalar fields, only the Dirac type Yukawa coupling 
with the type $\rho_2 \overline{n_L}{n_R} $ is allowed. 
All the other Yukawa couplings, such as $\rho_1 \overline{n_L}{n_R}$ ,  $\rho_i  \overline{(n_L)^c}{n_L}$ and 
 $\rho_i  \overline{(n_R)^c}{n_R}$ are forbidden 
by considering both $Z_{2}$ and $Z^\prime_{2}$ symmetries. 
About the neutrino mass term with the dimension three,  Dirac mass term $m_\nu \overline{n_L}{n_R}$ is forbidden due to $Z_{2}$ symmetry. To forbid Majorana
mass terms such as $\overline{(n_R)^c}{n_R}$ and $\overline{(n_L)^c}{n_L}$, we impose the symmetry under the transformation,  
$(n_L, n_R)$ into $ e^{i\frac{\pi}{2}} (n_L, n_R)$. 
\begin{table}[h]
	\begin{center}	
		\begin{tabular}{|c|c|c|c|c|} \hline \hline 
			Symmetry & $\rho_{1}$ & $\rho_{2}$ & $n_{L}$ & $n_{R}$  \\ \hline 
			$Z_2$ & -- & -- & + & -- \\ \hline 
			$Z_2'$ & -- & + & + & + \\ \hline \hline 
		\end{tabular}
\caption{The charge assignment under $Z_2$ and $Z_2'$ symmetries.}
\label{tab:tbA}
	\end{center}	
\end{table} 

Next we rewrite the action in Eq.($\ref{Action_bare}$) in terms of the renormalized fields, renormalized couplings and masses. 
The relations between bare quantities and renomalized ones are given as follows; 
\begin{eqnarray}
  \rho_{0 i} & = & \sqrt{Z_i} \rho_i ,\label{rho0i}\\
 n_{0} & =& \sqrt{Z_n} n, \\
  m_{0 i}^2 Z_i & = & \sum_{j=1}^2 Z_{m ij} m_j^2 , \label{eq:cmij} \\
  m_{012}^2 \sqrt{Z_1 Z_2} & = & m^2_{12} Z_{12} , \label{eq:cm12}\\
  \lambda_{0i} Z_i^2 & = & \sum_{I=1}^3 Z_{\lambda_{iI}} \lambda_I \mu^{2 \eta} , \label{eq:clami}\\
  \lambda_{03} Z_1 Z_2 & = &\sum_{I=1}^3 Z_{\lambda_{3I}} \lambda_I \mu^{2 \eta} 
,\label{eq:clam3} \\
  y_0 Z_n \sqrt{Z_2} &=& Z_y y \mu^\eta,
 \label{eq:cyukawa}
\end{eqnarray}
where the index $i$  ($i = 1,2$) is not summed, $\mu$ denotes the renormalization scale and $\eta$ is $2-\frac{d}{2}$.
As for the cosmological constants, the relations between the renormalized parameters and the bare ones are given as follows,
\bea
 \sum_{i=1}^{2} h_{0 i} m^4_{0 i} +  2 h_{03} m^2_{01} m^2_{02}
&=&\mu^{-2 \eta} \left(\sum_{i=1}^{2}Z_{h i} h_i {m_i}^4+ 
 2 Z_{h3} h_3 m^2_{1} m^2_{2} \right) , \label{rccos1} \\
 h_{0 12} {m_0}_{12}^4&=& \mu^{-2 \eta}  Z_{h12} h_{12} m_{12}^4. \label{rccos2}
\eea
 Using Eqs.\eqref{rho0i}-\eqref{rccos2}, the action in Eq.\eqref{Action_bare} can be written in terms of renormalized quantities as,
\begin{align}
  S[\rho_1,\rho_2,n]  = & - \frac{1}{2} \int d^d x \sum_{i=1}^2  \left( Z_i \rho_i \Box \rho_i  +\rho_i^2 Z_{m ij} m_j^2   
  	+ \frac{\mu^{2 \eta}}{2} \sum_{I=1}^3 \left( \rho_i^4  Z_{\lambda_{iI}} \lambda_I + \rho_1^2 \rho_2^2 Z_{\lambda_3I}\lambda_I \right)
\right) \nn \\
	& - \int d^d x \left( Z_y y \mu^\eta \bar{n} n  + Z_{12} m_{12}^2 \rho_1  \right) \rho_2
\nn  \\
& +\int d^d x \mu^{-2 \eta} \left(\sum_{i=1}^{2}Z_{h i} h_i {m_i}^4+ 
 2 Z_{h3} h_3 m^2_{1} m^2_{2} + Z_{h12} h_{12} m_{12}^4 \right).
 \label{action2}
\end{align}

\section{Definition of the effective action}
In this section, we give a definition of the effective action for
the light scalar $\rho_1$ by integrating the heavy scalar $\rho_2$ as well as the quantum fluctuation of the light scalar.  
To begin with, we  define the generating functional $W[J_1,n]$,
\begin{eqnarray}
  e^{i W [J_1, n]}  =  \int d \rho_1 \int d \Delta_2 e^{i S [\rho_1,
  \Delta_2, n] + i \int \rho_1 J_1 d^4 x} ,  \label{W1}
\end{eqnarray}
where we have introduced the source term $J_1$ for the lighter field $\rho_1$. About the heavy scalar field, we do not introduce 
any source term.  Instead, we integrate it by setting $\rho_2=\Delta_2$ in the action Eq.\eqref{action2} where we expand the heavy scalar field around the vanishing VEV.
 $\bar{\rho}_1$ which is the expectation value of $\rho_1$ is defined by,
\begin{eqnarray}
  \bar{\rho}_1 |_{J_1} \nobracket = \frac{\delta W [J_1, n]}{\delta J_1} =
  \frac{\int d \rho_1 \int d \Delta_2 \rho_1 e^{i S [\rho_1, \Delta_2, n] + i
  \int \rho_1 J_1 d^4 x}}{\int d \rho_1 \int d \Delta_2 e^{i S [\rho_1,
  \Delta_2] + i \int \rho_1 J_1 d^4 x}} \label{rho1bar}.
\end{eqnarray}
Then one can define the effective action $\Gamma_{\tmop{eff}}$, a functional of
$\bar{\rho}_1$ by Legendre transformation of $W [J_1, n] $ as,
\begin{align}
	\Gamma_{\tmop{eff}} [\bar{\rho}_1, n] & = W [J_1, n] - \int J_1 \bar{\rho}_1 d^4 x \notag \\
		 &=  - i \log \int d \Delta_1 \int d \Delta_2 e^{i S [\bar{\rho}_1 + \Delta_1, \Delta_2, n] 
		 	- i \int \Delta_1 \frac{\delta \Gamma_{\tmop{eff}} [\bar{\rho}_1, n]}{\delta \bar{\rho}_1 (x)} d^4 x} 	, \label{gamma3} \\
	& \Delta_1 \equiv \rho_1-\bar{\rho}_1 \label{Delta1},
\end{align}
where we substitute the relation $J_1 (x)=- \frac{\delta \Gamma_{\tmop{eff}} [\bar{\rho}_1, n]}{\delta \bar{\rho}_1 (x)}$. 
In  Eq.(\ref{gamma3}), 
We change the path integral variable from $\rho_1$ to $\Delta_1$ as defined in Eq.(\ref{Delta1}). $\Delta_1$ is  the quantum fluctuation from the expectation value $\bar{\rho}_1$ .
Using Eq.($\ref{rho1bar}$), one can show that,
\begin{eqnarray}
  \int d \Delta_1 \Delta_1 \int d \Delta_2 e^{i S [\bar{\rho}_1 + \Delta_1,
  \Delta_2, n] - i \int \Delta_1 \frac{\delta \Gamma_{\tmop{eff}}
  [\bar{\rho}_1, n]}{\delta \bar{\rho}_1 (x)} d^4 x} & = & 0 .
\end{eqnarray}
The tadpole vanishing condition for $\Delta_1$ leads to the one particle irreducibility
of the effective action
$\Gamma_{\tmop{eff}} [\bar{\rho}_1, n]$. Concerned with the heavier field $\Delta_2$,
the one particle reducible diagram is included. 
\section{Integrating the scalar fields}
In this section, we will integrate the scalar fields out.
We assume the following hierarchy for the mass parameters.
\bea
m_2^2 \gg - m_1^2 \simeq \epsilon m_{12}^2>0 ,\quad \epsilon = \frac{m_{12}^2}{m_2^2} \ll 1
\eea
We first integrate the heavy scalar field
$\Delta_2$  in Eq.(\ref{gamma3}).  Next, we integrate the lighter scalar field $\Delta_1$ 
which denotes the quantum fluctuation around the background field $\bar{\rho}_{1}$.  Concerning the corrections, we keep them up to the second order of the
coupling constants within 1 loop approximation. 
Below we show the steps to integrate the heavy scalar field $\Delta_2$ and 
the light scalar field $\Delta_1$.

\subsection{Integrating heavy scalar field}
Using the relation $\Delta_1$ in Eq.\eqref{Delta1} and Eq.\eqref{action2}, the action in
Eq.(\ref{gamma3}) is written as,
\begin{eqnarray}
  S [\bar{\rho}_1 + \Delta_1, \Delta_2, n] & = & S [\bar{\rho}_1, 0, 0] 
  + \int d^d x \Delta_i (x) \frac{\delta S [\rho_1, \rho_{2,} n]}{\delta \rho_i (x)}|_{\rho_1 = \bar{\rho}_1, \rho_2 = 0} \notag \\
  & + & \frac{1}{2} \int d^d x \int d^d y \Delta_i (x) \frac{\delta S [\rho_1, \rho_{2,} n]}{\delta \rho_i (x) \delta \rho_j (y)} 
  |_{\rho_1 = \bar{\rho}_1, \rho_2 = 0} \nobracket \Delta_j (y) \notag \\
  & +& S_{\tmop{int}} (\Delta_i, \bar{\rho}_1) \label{eq8}.
\end{eqnarray}
Each term in Eq(\ref{eq8}) is given as follows,
\begin{eqnarray}
	S [\bar{\rho}_1, 0, 0] & = & \int d^d x \left( \frac{Z_1}{2}\partial^{\mu} \bar{\rho}_1 \partial_{\mu} \bar{\rho}_1 
	  	- \frac{Z_{m 11}}{2}m_1^2 \bar{\rho}_1^2 - \frac{\lambda_{0 1} Z_1^2}{4}  \bar{\rho}_1^4 \right) \nn \\
	  	& + & \int d^d x \mu^{-2 \eta} \left(\sum_{i=1}^{2}Z_{h i} h_i {m_i}^4+ 
 2 Z_{h3} h_3 m^2_{1} m^2_{2} + Z_{h12} h_{12}m_{12}^4 \right)\\
	\frac{\delta S [\rho_1, \rho_{2,} n]}{\delta \rho_i (x)} |_{\rho_1 =
  \bar{\rho}_1, \rho_2 = 0} \nobracket & =& - \left(\begin{array}{c}
    \{ Z_1 (\Box + m_{01}^2) + \lambda_{01} Z_1^2  \bar{\rho}_1^2 \}
    \bar{\rho}_1 \quad\\
    \mu^{\eta} y Z_y  \bar{n} n + Z_{12} m_{12}^2 \bar{\rho}_1
  \end{array}\right), \\
    \frac{\delta S [\rho_1, \rho_{2,} n]}{\delta \rho_i (x) \delta \rho_j (y)}
  |_{\rho_1 = \bar{\rho}_1, \rho_2 = 0} \nobracket & =& -
  \left(\begin{array}{cc}
    Z_1 (\Box + m_{01}^2) + 3 \lambda_{01} Z_1^2  \bar{\rho}_1^2  & Z_{12}
    m_{12}^2\\
    Z_{12} m_{12}^2 & Z_2 (\Box_x + m_{02}^2) + \frac{\lambda_{0 3}}{2} Z_1
    Z_2 \bar{\rho}_1^2
  \end{array}\right)  \notag \\ && \times \delta^d (x - y), \\
	S_{\tmop{int}} (\Delta_i, \bar{\rho}_1) & = & S_{1 \tmop{int}} (\Delta_1, \bar{\rho}_1) + S_{2 \tmop{int}} (\Delta_2) + S_{12 \tmop{int}}
		(\Delta_1, \Delta_2, \bar{\rho}_1) , \\ 
	S_{1 \tmop{int}} (\Delta_1, \bar{\rho}_1) & = & - \int d^d x \lambda_{01} Z_1^2 \Delta_1^3 \bar{\rho}_1  
		- \int d^d x \frac{1}{4} \lambda_{01} Z_1^2 \Delta_1^4 ,\\
	S_{2 \tmop{int}} (\Delta_2) & = & - \int d^d x \frac{1}{4} \lambda_{02} Z_2^2 \Delta_2^4 , \\
	S_{12 \tmop{int}} (\Delta_1, \Delta_2, \bar{\rho}_1) & = & - \int d^d x \frac{1}{2} \lambda_{03} Z_1 Z_2 \Delta_1 \Delta_2^2 \bar{\rho}_1 
	- \int d^d x \frac{1}{4} \lambda_{03} Z_1 Z_2 \Delta_1^2 \Delta_2^2.
\end{eqnarray} 
We define the following quantity by subtracting the classical action from the effective action,
\begin{eqnarray}
  \tilde{\Gamma}_{\tmop{eff}} [\bar{\rho}_1, n] & = & \Gamma_{\tmop{eff}}
  [\bar{\rho}_1, n] - S [\bar{\rho}_1, 0, 0]  \label{defgammatilde},
\end{eqnarray}
with $\tilde{\Gamma}_{\tmop{eff}} [\bar{\rho}_1, n]$ is written as follows,
\begin{align}
  e^{i \tilde{\Gamma}_{\tmop{eff}} [\bar{\rho}_1, n]} & =  \int d \Delta_1
  e^{i \left\{ \frac{1}{2} \int d^d x \int d^d y \Delta_1 (x) \frac{\delta^2 S
  [\bar{\rho}_1, 0, n]}{\delta \bar{\rho}_1 (x) \delta \bar{\rho}_1 (y)}
  \Delta_1 (y) + S_{1 \tmop{int}} (\Delta_1, \bar{\rho}_1) - \int d^d x
  \Delta_1 (x) \left( \frac{\delta \tilde{\Gamma}_{\tmop{eff}} [\bar{\rho}_1,
  n]}{\delta \bar{\rho}_1 (x)} \right) \right\}} e^{i W_2 [J_2, \Delta_1, n]}. \label{eq17}
\end{align}
The last factor of Eq(\ref{eq17}) summarizes the contribution from the heavy field to the effective action.
It is written in terms of the path integral with respect to $\Delta_2$ as follows,
\begin{align}
	& e^{i W_2 [J_2,\bar{\rho}_1, \Delta_1]}  \nonumber \\
		 = & \int d \Delta_2 e^{\frac{i}{2} \int d^d x \int d^d y \Delta_2 (x) 
			\frac{\delta^2 S [\rho_1, \rho_{2,} n]}{\delta \rho_2 (x) \delta \rho_2 (y)} |_{\rho_1 = \bar{\rho}_1, \rho_2 = 0} \Delta_2 (y) 
		+ i S_{2 \tmop{int}} (\Delta_2) + i S_{12 \tmop{int}} (\Delta_1, \Delta_2, \bar{\rho}_1) 
		+ i \int d^d x \Delta_2 (x) J_2 (x)}  \label{defW2},
\end{align} 
where $J_2$ is given by,
\begin{align}
J_2 (x) = \left( \frac{\delta S [\rho_1, \rho_{2,}
n]}{\delta \rho_2 (x)} |_{\rho_1 = \bar{\rho}_1, \rho_2 = 0}  -
Z_{12} m_{12}^2 \Delta_1 \right) = - Z_{12} m^2_{12} (\Delta_1 + \bar{\rho}_1)
- \mu^{\eta} y Z_y  \bar{n} n.
\label{source}
\end{align}
$J_2$ consists of the fields which linearly couple to the heavy scalar $\Delta_2$.
We note that  $W_2$ in Eq.(\ref{defW2})  is the generating functional for 
Green functions for the heavy scalar $\Delta_2$. 
We calculate $W_2[J_2,\bar{\rho}_1, \Delta_1]$ by integrating the heavy scalar field $\Delta_2$. 
The result can be compactly written by introducing  the exponentiated functional differentiation and the notation $\langle \cdot \cdot \cdot \rangle_0$ as,
\bea
	\braket{F[\Delta_2]}_0\equiv \exp \left[\frac{1}{2} \frac{\delta}{\delta \Delta_2}\cdot D_{F 22 } \frac{\delta}{\delta \Delta_2 } 
\right]
		F[\Delta_2]\Big|_{\Delta_2=0}, 
\label{eq:F}
\eea
where
\bea
\frac{\delta}{\delta \Delta_2}\cdot D_{F 22 } \frac{\delta}{\delta \Delta_2} \equiv \int d^dx  \int d^d y \frac{\delta}{\delta \Delta_2(x)} D_{F 22 }(x,y) \frac{\delta}{\delta \Delta_2 (y)},
\eea
$D_{F ii }(x,y)$ denotes the Feynman propagator
for $\Delta_i$. Using the notation,  the generating functional is given as follows;
\begin{align}
	e^{i W_2 [J_2, \bar{\rho}_1, \Delta_1]} 
		& = \left( \det \frac{\delta^2 S [\rho_1, \rho_{2,} n]}{\delta \rho_2 (x) \delta \rho_2 (y)} 
		\Bigr{|}_{\rho_1 = \bar{\rho}_1, \rho_2 = 0} \right)^{- \frac{1}{2}} \langle 
		e^{i S_{2 \tmop{int}} (\Delta_2) + i S_{12 \tmop{int}} (\Delta_1, \Delta_2, \bar{\rho}_1)} \rangle_0 \ 
		e^{i W_2^c [J_2,  \bar{\rho}_1,\Delta_1]} , \label{AfterW2}
\end{align}
where $e^{i W_2^c [J_2,\bar{\rho}_1, \Delta_1]}$ is given as,
\begin{eqnarray}
  e^{i W_2^c [J_2,\bar{\rho}_1, \Delta_1]} & = &
  \frac{\left\langle e^{i S_{2 \tmop{int}} (\Delta_2) + i S_{12 \tmop{int}}
  (\Delta_1, \Delta_2, \bar{\rho}_1^{}) + i \int d^d x \Delta_2 (x) J_2 (x)}
  \right\rangle_0}{\langle e^{i S_{2 \tmop{int}} (\Delta_2) + i S_{12
  \tmop{int}} (\Delta_1, \Delta_2, \bar{\rho}_1^{})} \rangle_0}   \label{W2c}.
\end{eqnarray}
In Eq.(\ref{AfterW2}), the first factor is the vacuum graph contribution from the quadratic  part of $\Delta_2$
and the second part corresponds to that from the interaction. The third factor is the connected Green function contribution of
$\Delta_2$ with the source term $J_2$ as given in Eq.(\ref{source}).
Next we define $D^{-1}_{11}(x,y)$ and $D^{-1}_{22}(x,y)$ as follows,
\bea
   &&- D^{- 1}_{11}  (x, y)\equiv - \{ Z_1 (\Box_x + m_{01}^2) + 3 \lambda_{01} Z_1^2  \bar{\rho}_1^2  (x) \}
  \delta^d (x - y) , \\
  &&- D^{- 1}_{22} 
  (x, y) \equiv
- \left\{ Z_2 (\Box_x + m_{02}^2) + \frac{\lambda_{0 3}}{2} Z_1 Z_2
  \bar{\rho}_1^2 (x) \right\} \delta^d (x - y),
\eea
where $D^{- 1}_{ii}  (x, y)$ depends on the background field $\bar{\rho}_1$.
The propagators $D_{F11}(x,y)$ and $D_{F22}(x,y)$ are determined so that they satisfy the following equations,
\begin{eqnarray}
  \int d^dy i  D^{- 1}_{ii}(x,y) D_{F ii} (y,z) & = & \delta^d (x - z) \ \ (i=1,2).
\end{eqnarray}
The propagators are symbolically written as follows,
\begin{eqnarray}
  D_{F 11} (x, y) & = & - \frac{i}{Z_1 (\Box_x + m_{01}^2) + 3 \lambda_{01} Z_1^2  \bar{\rho}_1^2  (x)}, \\
  D_{F 22} (x, y) & = & - \frac{i}{Z_2 (\Box_x + m_{02}^2) +
  \frac{\lambda_{0 3}}{2} Z_1 Z_2 \bar{\rho}_1^2 (x)}. \label{Propa2v1}
\end{eqnarray}
Note that  $D_{Fii}(x,y)$ also depends on the background field $\bar{\rho}_1$.
One can write  $W_2^c [J_2, \Delta_1]$ as the sum of the contribution from the tree 
diagrams and that of one loop diagrams,
\begin{align}
i W_2^c[J_2, \bar{\rho}_1, \Delta_1] & = 
i W_2^{\tmop{c (tree)}}[J_2, \bar{\rho}_1, \Delta_1]  + i \bar{W_2}^{c (1 \tmop{loop})}[\bar{\rho}_1, n].
\label{W2csum}
\end{align}
$i W_2^c$ is obtained from the definition of Eq.(\ref{W2c}) by using the differentiation in Eq.(\ref{eq:F}).
In Eq.(\ref{W2csum}),   $i W_2^{\tmop{c (tree)}} $ is given as,
\begin{align}
	& i W_2^{\tmop{c(tree)}} [J_2, \bar{\rho}_1, \Delta_1] \nn \\
		&=  - \frac{1}{2}  \int d^d x \int d^d y J_2(x) D_{F 22} (x, y) J_2 (y)
			-  i \frac{\lambda_{02}}{4} \int d^d x  D_J(x)^4 \nn \\
			& -  i \frac{\lambda_3  \mu^{2\eta}}{4}  \int d^d x (\Delta_1^2 (x) + 2 \Delta_1 (x) \bar{\rho}_1 (x)) D_J(x)^2
			-  \frac{\lambda_{02}^2  }{2}  \int d^d x \int d^d y D_{F 22} (x, y) D_J(x)^3 D_J (y)^3 \notag \\
			&  - \frac{\lambda_3^2  \mu^{4\eta} }{8} \int d^d x \int d^d y (\Delta_1^2 (x) + 2 \Delta_1 (x) \bar{\rho}_1 (x)) 
			(\Delta_1^2 (y) + 2 \Delta_1 (y) \bar{\rho}_1 (y))   D_{F 22} (x, y) D_J (x) D_J (y) \notag \\
			& -  \frac{\lambda_2 \lambda_3  \mu^{4\eta}}{2} \int d^d x \int d^d y (\Delta_1^2 (x)
			+ 2 \Delta_1 (x) \bar{\rho}_1 (x)) D_{F22}(x,y) D_J(x) D_J (y)^3,
\label{W2tree1}
\end{align} 
where we have used the following definition,
\begin{align}
D_J (x)^n = \prod_{i=1}^n \int d^d x_i D_{F22}(x,x_i) i J_2(x_i), \quad (n=1,2,3...). \label{defDJ}
\end{align}
\begin{figure}
\begin{center}
\begin{tabular}{ccc}
\includegraphics[width=4.0cm]{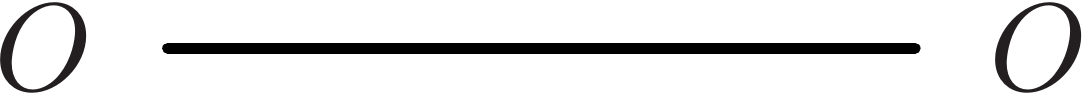} & \includegraphics[width=4.5cm]{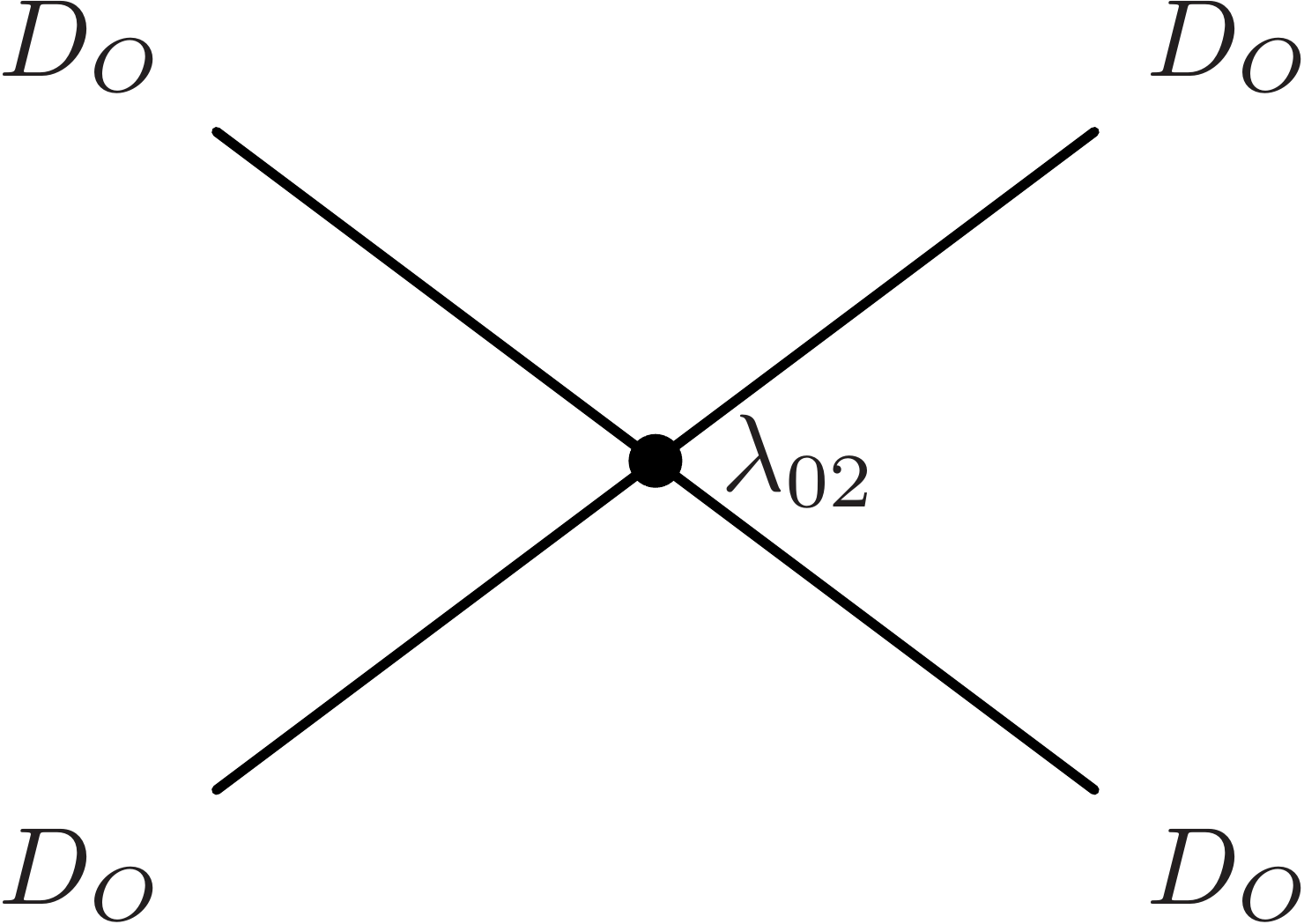}& \includegraphics[width=5.5cm]{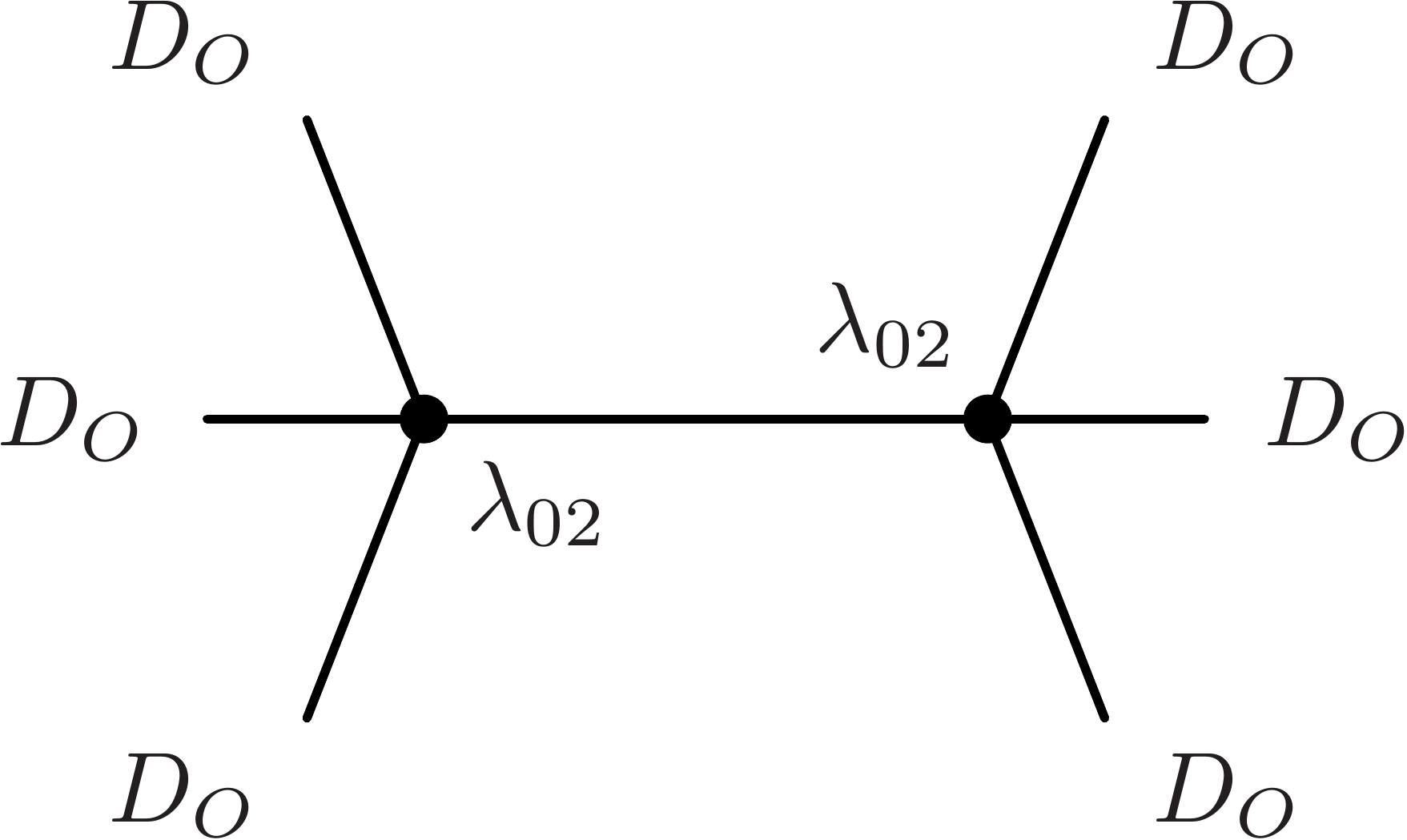} \\
\end{tabular}
\end{center}
\caption{The tree level diagrams which contribute to $i \bar{W}_2^{c (\tmop{tree})}[\bar{\rho}_1,n]$. All the propagators are heavy scalars and the external source terms
denoted by $O$ 
correspond to the background field $\bar{\rho}_1$ and the neutrino bilinear $\overline{n}n$.
 From the figure of the left to that of the right, each Feynman diagram corresponds to the terms of Eq.\eqref{W2tree1p} from the first term to the last term respectively. }\label{fig:fig1}
\end{figure}
The tree contribution is obtained by setting $\Delta_1$ to zero in
 Eq.(\ref{W2tree1}),
\bea
 i \bar{W}_2^{c (\tmop{tree})} [\bar{\rho}_1,n ]
	& \equiv&  i W_2^{c (\tmop{tree})} [J_2 (\Delta_1 = 0), \bar{\rho}_1,0] \nn \\
	& = & - \frac{1}{2}  \int d^d x \int d^d y O(x) D_{F 22} (x, y) O (y) 
	- i \frac{\lambda_{02}}{4} \int d^d x  D_O(x)^4   \notag \\
	&  &-    \frac{\lambda_{02}^2  }{2}  \int d^d x \int d^d y D_{F 22} (x, y) D_O(x)^3 D_O(y)^3,
\label{W2tree1p}
\eea
where we have defined $D_O (x)$ as follows,
\begin{align}
D_O (x) = \int d^d x_i D_{F22}(x,x_i) iO(x_i) , \label{defDO}
\end{align}
where  $-J_2(\Delta_1=0)=O(x)= Z_{12} m^2_{12} \bar{\rho}_1 + \mu^{\eta} y Z_y  \bar{n} n$.
The diagrams for Eq.(\ref{W2tree1p}) are shown in Fig.\ref{fig:fig1}.
All the propagators in the diagrams are the heavy scalars and they are one particle
reducible diagrams.
In Eq.(\ref{W2tree1p}), the bare  coupling constant $\lambda_{02}$ is substituted
because these tree level contribution includes the counter terms which subtract the divergences of one-loop graphs.  The terms with $\Delta_1$ in
 Eq.(\ref{W2tree1})  contribute to
the effective action beyond the tree level and the renormalized coupling constants
are substituted for their interactions.

\begin{figure}
\begin{center}
\begin{tabular}{ccc}
\includegraphics[width=5.0cm]{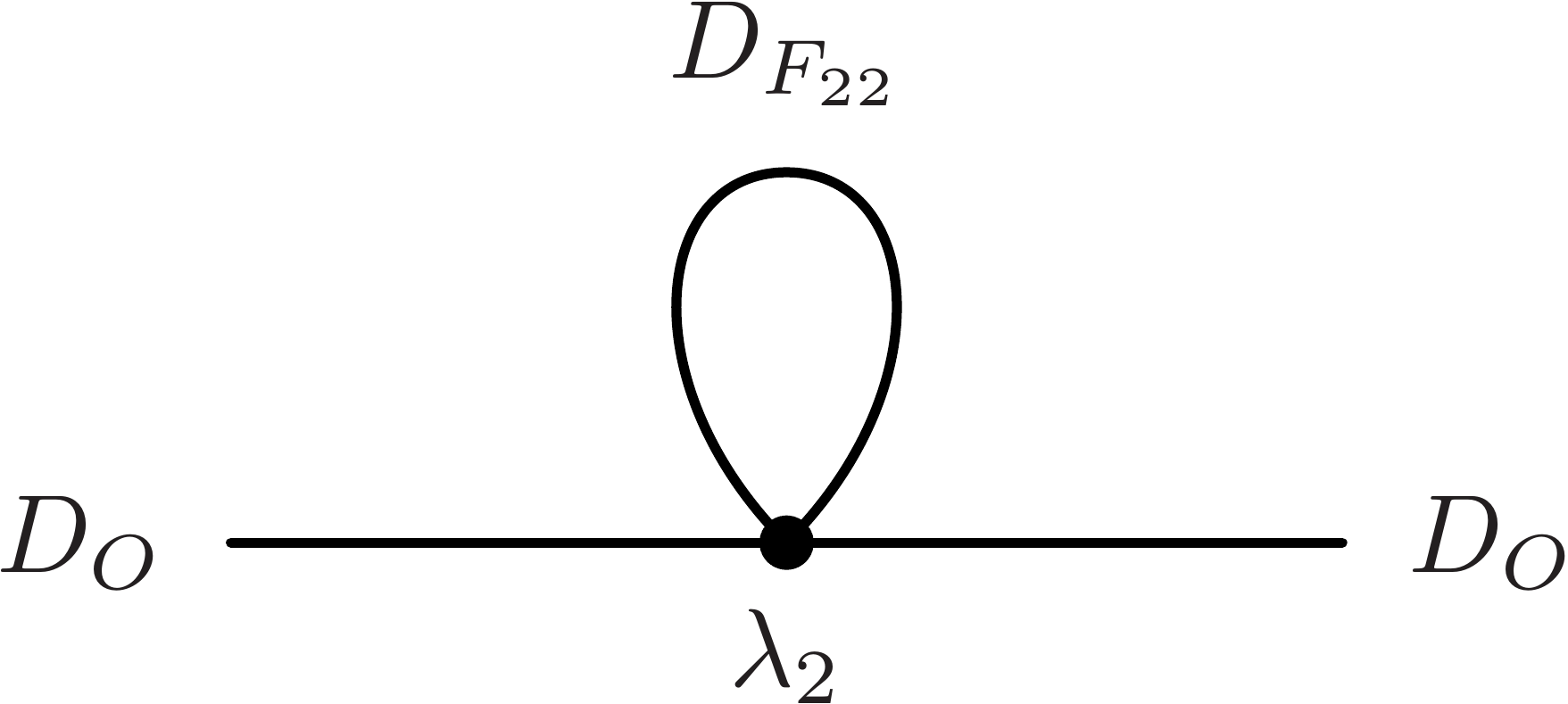} & \includegraphics[width=5.0cm]{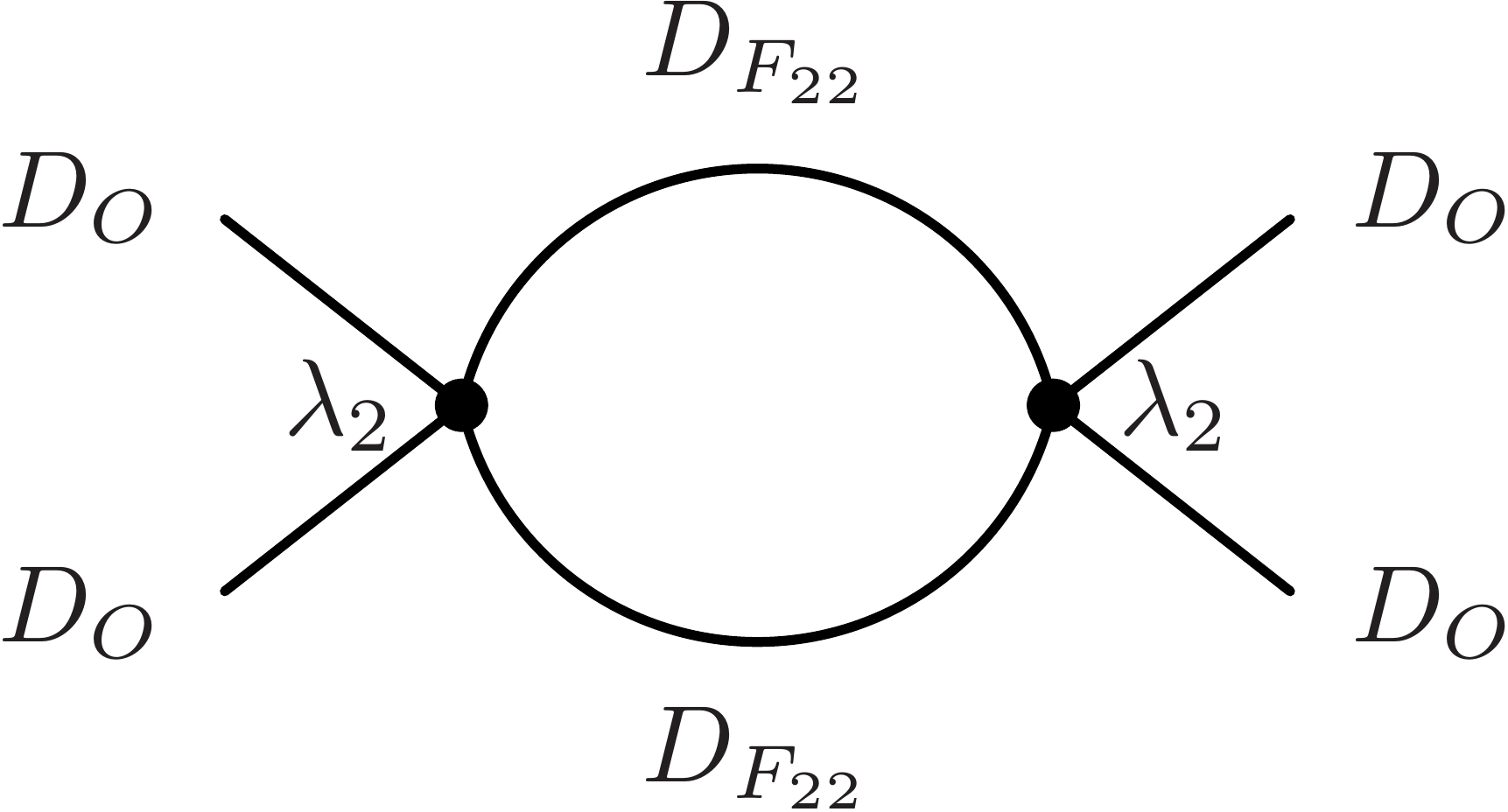}& \includegraphics[width=6.5cm]{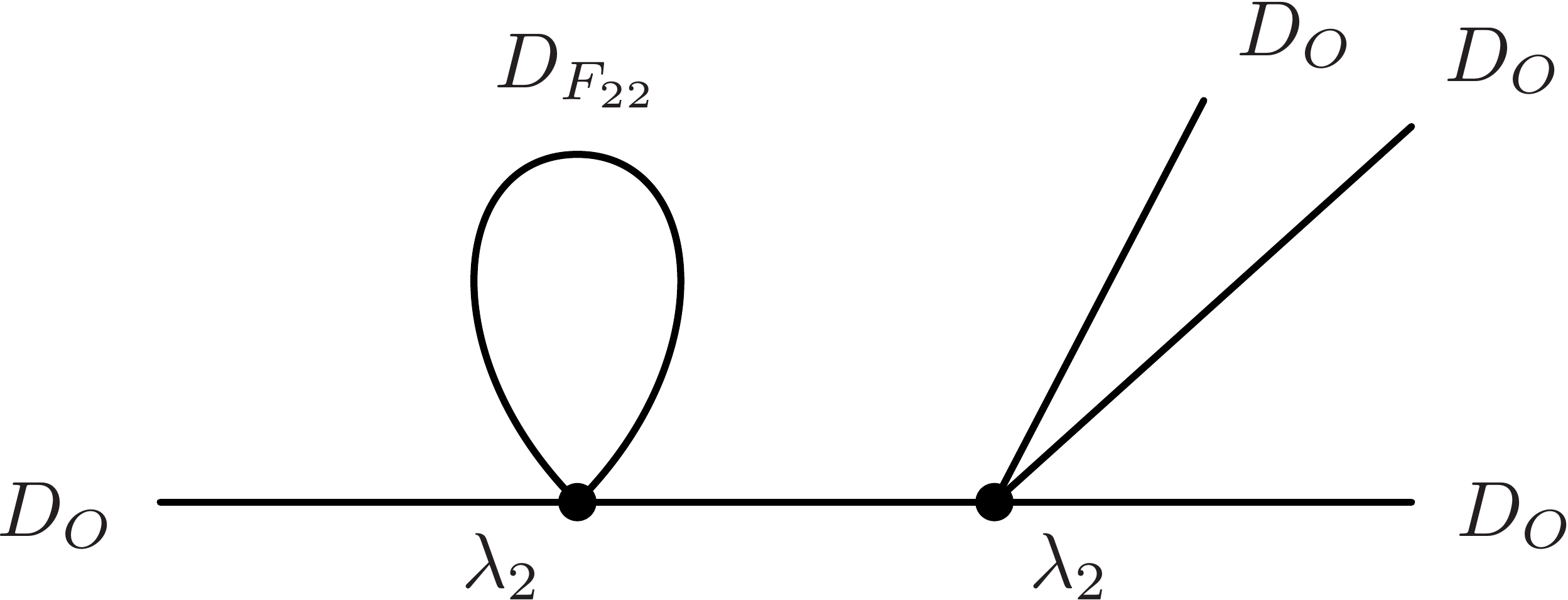} \\
\end{tabular}
\end{center}
\caption{One loop diagrams of the heavy scalar $\Delta_2$ which contribute to the  $i \bar{W}_2^{c (1\tmop{loop})} [\bar{\rho}_1,n ]$.
From the figure of the left to that of the right, each Feynman diagram corresponds to terms of Eq.\eqref{W2c1loop} from the first term to the last term respectively.}
\label{fig:fig2}  
\end{figure}

In Fig.\ref{fig:fig2}, we have shown the diagrams for
the one-loop contribution of $\Delta_2$ defined by $i \bar{W}_2^{c( 1 \tmop{loop})}$. The contribution is summarized as,
\begin{align}
     & i \bar{W}_2^{c (1\tmop{loop})} [\bar{\rho}_1,n ] 
	\equiv  i W_2^{c (1\tmop{loop})} [- O (x), \bar{\rho}_1,0] \nn \\
	&	=  -\frac{3 i \lambda_2 \mu^{2\eta} }{2}   \int d^d x D_{F 22} (x, x)  D_O(x)^2 \nn \\
			& - \frac{\lambda_2^2  \mu^{4\eta} }{2} \int d^d x \int d^d y 
			\left(\frac{9}{2}  D_{F 22} (x, y)^2 D_O(x)^2 D_O(y)^2+ 6D_{F 22} (x,y)D_{F22} 
(x,x)D_O(x) D_O(y)^3 \right). 
\label{W2c1loop}
\end{align}
In this contribution, one sets $\Delta_1$ equal to zero,
since the contribution of $\Delta_1$ generates either another loop effect or one-particle reducible contribution  which is excluded from $\tilde{\Gamma}_{\mathrm{eff}}[\bar{\rho}_1,n]$ in Eq.(\ref{eq17}). With  Eq.(\ref{W2tree1}) and Eq.(\ref{W2c1loop}) , the calculation of $W_2^c$ is completed.  

Next we move to the vacuum graph contribution; i.e., the second factor of 
Eq.(\ref{AfterW2}) and it is computed as,
\begin{align}
	&\langle e^{i S_{2 \tmop{int}} (\Delta_2) + i S_{12 \tmop{int}} (\Delta_1, \Delta_2, \bar{\rho}_1)} \rangle_0 \notag \\
	& =  \exp \left[- \frac{\lambda_3  \mu^{2\eta} }{4} i \int d^d x (\Delta_1^2 (x) + 2 \Delta_1 (x) \bar{\rho}_1 (x)) D_{F 22} (x, x) \right. \notag \\
	& \left. - \left( \frac{\lambda_3}{4}  \right)^2  \mu^{4\eta} \int d^d x \int d^d y (\Delta_1^2 (x) + 2 \Delta_1 (x) \bar{\rho}_1 (x)) 
		 (\Delta_1^2 (y) + 2 \Delta_1 (y) \bar{\rho}_1 (y)) D^2_{F 22} (x, y) \right] \nn \\
&=1 
\label{Lvac}.
\end{align}
We note in the contribution, one loop contribution of heavy scalar is present. Therefore  another loop effect of the light scalar leads to
two loop contribution. 
From the same reason as that of $i W_2^{c (1 \tmop{loop}) }$,
we set $\Delta_1$ to be zero in the last line of Eq.(\ref{Lvac}). 
Combining Eq.(\ref{W2tree1}) and Eq.(\ref{W2c1loop}), one can summarize the expression for Eq.(\ref{defW2}).
\begin{align}
	e^{i W_2 [J_2, \bar{\rho}_1, \Delta_1]} 
		=e^{-\frac{1}{2} \tmop{TrLn} D^{- 1}_{22} (\bar{\rho}_1)+i \bar{W}_2^{c (\tmop{tree})} [\bar{\rho}_1,n ]
		+i \bar{W}_2^{c (1\tmop{loop})} [\bar{\rho}_1,n ] }
		e^{i W_2^{c (\tmop{tree})}[J_2, \bar{\rho}_1, \Delta_1]|_{\mathrm{rest}} } , \label{AAfterW2}
\end{align}
where $i W_2^{c (\tmop{tree})}|_{\mathrm{rest}}$ is defined as the difference  of Eq.(\ref{W2tree1}) and Eq.(\ref{W2tree1p}),
\bea
	i W_2^{c (\tmop{tree})}[J_2, \bar{\rho}_1, \Delta_1]|_{\mathrm{rest}}
		\equiv i W_2^{\tmop{c(tree)}} [J_2, \bar{\rho}_1, \Delta_1]
		-i \bar{W}_2^{c (\tmop{tree})} [\bar{\rho}_1,n ] .
\eea

\subsection{Integrating lighter scalar field}
In the following, we integrate the lighter scalar field $\Delta_1$.  We include the corrections up to one loop level.  
Using the result of previous section,  one loop part of the effective action is written by,
\bea
  && e^{i \tilde{\Gamma}_{\tmop{eff}} [\bar{\rho}_1, n] }= 
   e^{- \frac{1}{2} \tmop{TrLn} D^{- 1}_{11} (\bar{\rho}_1)} 
	 e^{-\frac{1}{2} \tmop{Tr} \tmop{Ln} D_{22}^{- 1} (\bar{\rho}_1)} 
	e^{i \bar{W}_2^{c (\tmop{tree})} [\bar{\rho}_1,n ]
		+i \bar{W}_2^{c (1\tmop{loop})} [\bar{\rho}_1,n ] } \notag \\
  && \times \frac{\int d \Delta_1 e^{i \left\{ \frac{1}{2} \int d^d x \int d^d y
  \Delta_1 (x) \frac{\delta^2 S [\bar{\rho}_1, 0, n]}{\delta \bar{\rho}_1 (x)
  \delta \bar{\rho}_1 (y)} \Delta_1 (y) + S_{1 \tmop{int}} \left( \Delta_1,
  \bar{\rho}_1 \right) + W_2^{c (\tmop{tree})}[J_2, \bar{\rho}_1, \Delta_1]|_{\mathrm{rest}}
  - \int d^d
  x \Delta_1 (x) \left( \frac{\delta \tilde{\Gamma}_{\tmop{eff}}
  [\bar{\rho}_1, n]}{\delta \bar{\rho}_1 (x)} \right) \right\}}_{} 
 }{\int d \Delta_1 e^{i \left\{ \frac{1}{2} \int
  d^d x \int d^d y \Delta_1 (x) \frac{\delta^2 S [\bar{\rho}_1, 0, n]}{\delta
  \bar{\rho}_1 (x) \delta \bar{\rho}_1 (y)} \Delta_1 (y) \right\}}}. \nn \\
\label{eq:eiGamma}
\eea
 From Eq.(\ref{eq:eiGamma}), one obtains the effective action as. 
\begin{eqnarray}
  i \tilde{\Gamma}_{\tmop{eff}} [\bar{\rho}_1, n] & = & - \frac{1}{2}
  \tmop{TrLn} D^{- 1}_{22} (\bar{\rho}_1) - \frac{1}{2} \tmop{Tr} \tmop{Ln}
   \{ (D_{11}^{- 1})' (\bar{\rho}_1)  \} + i \bar{W}_2^{c (\tmop{tree})} [\bar{\rho}_1,n ]
		+i \bar{W}_2^{c (1\tmop{loop})} [\bar{\rho}_1,n ]  \nn \\
  & + & \log \left[ \frac{\int d \Delta_1 e^{i \left\{ -\frac{1}{2} \int d^d x
  \int d^d y \Delta_1 (x) \{  (D_{11}^{- 1})' (\bar{\rho}_1)
\} \Delta_1 (y) - \int d^d x \Delta_1
  (x) \left( \frac{\delta \tilde{\Gamma}_{\tmop{eff}} [\bar{\rho}_1,
  n]}{\delta \bar{\rho}_1 (x)} \right) \right\}}_{} e^{\mathcal{L}_{\tmop{int}} (\Delta_1)} }{\int d \Delta_1 e^{i \left\{ - \frac{1}{2} \int d^d x
  \int d^d y \Delta_1 (x) \{  (D_{11}^{- 1})' (\bar{\rho}_1)
\}     \Delta_1 (y) \right\}}_{}} \right]. \nn \\
\label{eq: iGamma}
\end{eqnarray}
In Eq.(\ref{eq: iGamma}), we define $\mathcal{L}_{\tmop{int}} (\Delta_1) $ by;
\bea
&&
\mathcal{L}_{\tmop{int}} (\Delta_1) \nn \\
&& \equiv i S_{1 \tmop{int}} (\Delta_1, \bar{\rho}_1) + i W_2^{c \tmop{tree}} [J_2, \Delta_1, \bar{\rho}_1 ]|_{\mathrm{rest}}
 +   \frac{1}{2} m_{12}^4 \int d^d x \int d^d y \Delta_1 (x) D_{F 22} (x, y) \Delta_1 (y) \nn \\
&&\simeq  i W_2^{c \tmop{tree}}[J_2, \Delta_1, \bar{\rho}_1 ]|_{\mathrm{rest}}^{\mbox{\tiny quadratic part of $\Delta_1$}}+  \frac{1}{2} m_{12}^4 \int d^d x \int d^d y \Delta_1 (x) D_{F 22} (x, y) \Delta_1 (y) ,\nn \\
\label{eq:Lint}
\eea
To derive Eq.(\ref{eq: iGamma})  from Eq.(\ref{eq:eiGamma}),  we absorb
the mixing effect of the heavy scalar into the propagator of  the light scalar.  
The corresponding inverse propagator is given as,  
\bea
  - D_{11}^{- 1'}(x,y)  =  - D^{- 1}_{11}  (x, y) + i m_{12}^4 D_{F 22} (x, y). 
\label{eq:newprop}
\eea
The mixing of the heavy scalar  is the second term of Eq.(\ref{eq:newprop})  and the modified propagator for the light scalar is defined by the following equation,
\begin{align}
i \int d^d z  D_{11}^{- 1 '} (x,z) D'_{F 11} (z, y) =  \delta^d (x - y).
\end{align}
Because of this change,  the quadratic term with respect to $\Delta_1$ in 
$ i W_2^{c \tmop{tree}} [J_2, \Delta_1, \bar{\rho}_1 ]|_{\mathrm{rest}}$;
\bea
-\frac{1}{2}m_{12}^4 \int \int d^{d}{x} d^{d}{y}\Delta_1(x) D_{F22}(x,y) \Delta_1(y),
\eea
should be subtracted. 
The  second term of the last line of  Eq.(\ref{eq:Lint}) is added for this purpose.
As for the other parts of Eq.(\ref{eq:Lint}), 
from the second line to 
the third line of Eq.(\ref{eq:Lint}), the quartic interaction term with respect to $\Delta_1$ is ignored because it contributes to beyond the one loop order.
The contribution to the tadpole diagram from the cubic interaction  of $\Delta_1$ is absent in 1  PI effective action and the second order contribution from the cubic interaction 
is also ignored because  it contributes to beyond the one loop order.  Therefore 
within one-loop approximation , we keep only the quadratic terms with respect to $\Delta_1$. The explicit expression is given as follows,
\footnotesize
\begin{align}
& \mathcal{L}_{\tmop{int}} (\Delta_1) \equiv 
		- \frac{3 i m_{12}^4 \lambda_2 \mu^{2\eta} }{2}  \int d^d x 
			\left( \prod_{i = 1}^4 \int d^d x_i D_{F22} (x,x_i) \right) O (x_3) O (x_4) \Delta_1 (x_1) \Delta_1 (x_2) \notag \\
			& +  \frac{i \lambda_3\mu^{2\eta}}{4} \int d^d x \Delta_1^2 (x) 
			\left( \prod_{i = 1}^2 \int d^d x_i D_{F22} (x,x_i) O(x_i)\right) \notag \\
			& + i m_{12}^2 \lambda_3 \mu^{2\eta} \int d^d x \Delta_1 (x) \bar{\rho}_1(x) 
			\left( \prod_{i = 1}^2 \int d^d x_i D_{F22} (x,x_i) \right) \Delta_1 (x_1) O (x_2) \notag \\
			& +  \frac{\lambda_3^2\mu^{4\eta}}{2} \int d^d x \int d^d y \Delta_1 (x) \bar{\rho}_1 (x) \Delta_1 (y) \bar{\rho}_1(y) 
			D_{F 22} (x, y)  \int d^d x_1 \int d^d y_1 D_{F 22} (x, x_1) D_{F 22} (y, y_1) O (x_1) O (y_1) \notag \\
			& -  \frac{\lambda_2 \lambda_3 \mu^{4\eta} }{4}  \int d^d x \int d^d y \Delta_1^2 (x) D_{F 22} (x, y)
			\int d^dx_1 D_{F 22} (x, x_1)O(x_1) \left( \prod_{i = 1}^3 \int d^d y_i D_{F 22} (y, y_i) O (y_i) \right) \notag \\
			& -   \frac{m_{12}^2 \lambda_2 \lambda_3 \mu^{4\eta}}{2}  \int d^d x \int d^d y \Delta_1 (x) \bar{\rho}_1 (x) D_{F 22} (x, y)
			\int d^dx_1 D_{F 22} (x, x_1) \Delta_1 (x_1)\left( \prod_{i = 1}^3 \int d^d y_i D_{F 22} (y, y_i) O (y_i)\right)
			\notag \\
			& -  \frac{3}{2} m_{12}^2 \lambda_2 \lambda_3 \mu^{4\eta} \int d^d x \int d^d y \Delta_1 (x) \bar{\rho}_1 (x) D_{F 22} (x, y)
			\int d^dx_1 D_{F 22} (x, x_1) O (x_1) \left( \prod_{i = 1}^3 \int d^d y_i D_{F22} (y,y_i)\right) \Delta_1(y_1) O(y_2)O(y_3) 
			\notag \\
			& +  3  m_{12}^4 \lambda_2^2 \mu^{4\eta} \int d^d x \int d^d y D_{F 22} (x, y) 
			\left( \prod_{i = 1}^3 \int d^d x_i D_{F22} (x,x_i) \right)\Delta_1(x_1) \Delta_1(x_2) O(x_3) 
			\left( \prod_{j = 1}^3 \int d^d y_j D_{F22} (y,y_j) O(y_j)\right)  \notag \\
			& +   \frac{9 m_{12}^4 \lambda_2^2 \mu^{4\eta}}{2} \int d^d x \int d^d y D_{F 22} (x, y) 
			\left( \prod_{i = 1}^3 \int d^d x_i \int d^d y_i D_{F22} (x,x_i) D_{F22} (y,y_j)  \right)
			\Delta_1(x_1) O(x_2) O(x_3) \Delta_1(y_1) O(y_2) O(y_3) .
\label{eq:lint}
\end{align}
\normalsize
\begin{figure}
\begin{center}
\begin{tabular}{ccc}
\includegraphics[height=2.25cm]{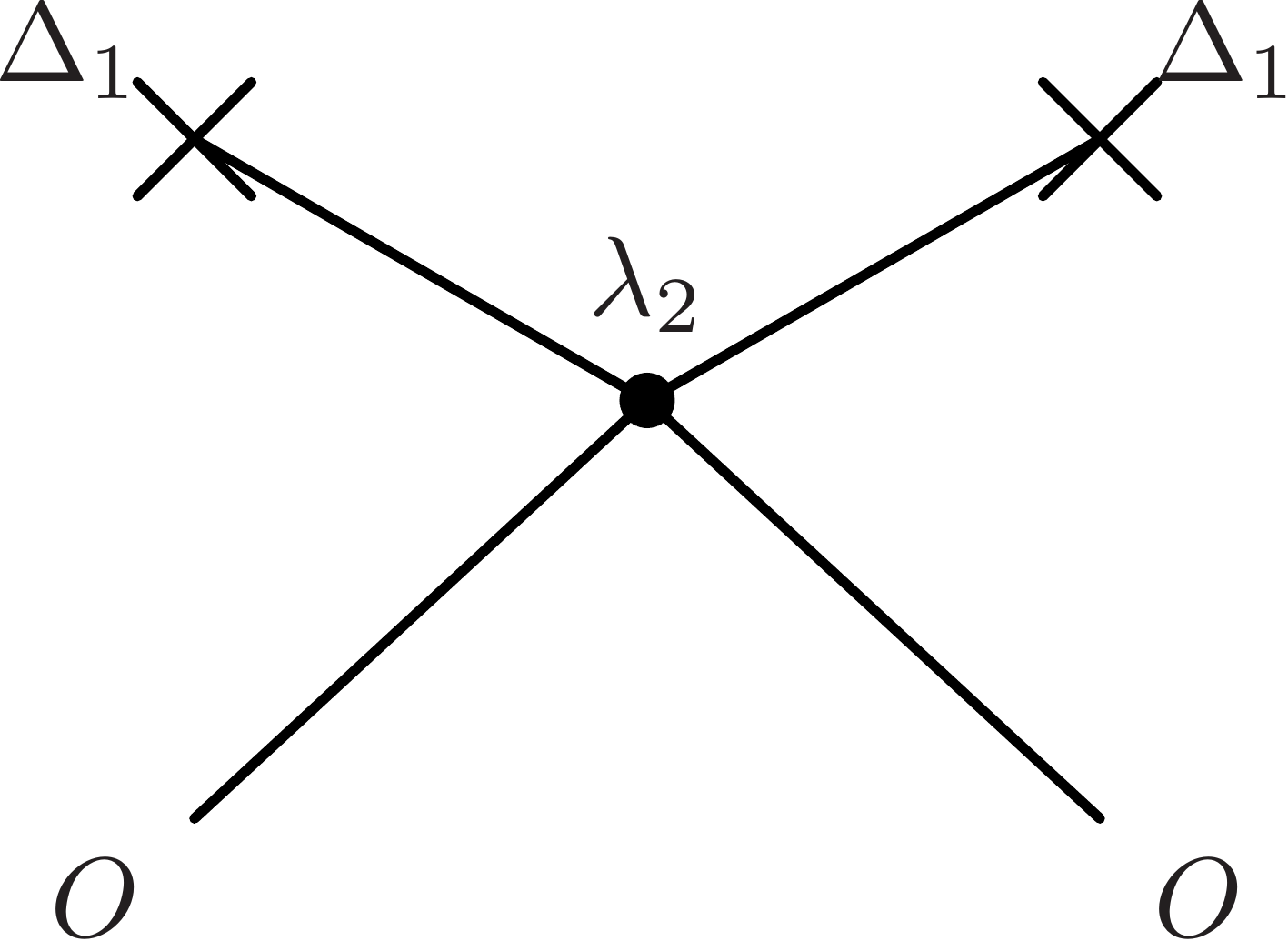} & \includegraphics[width=4.5cm]{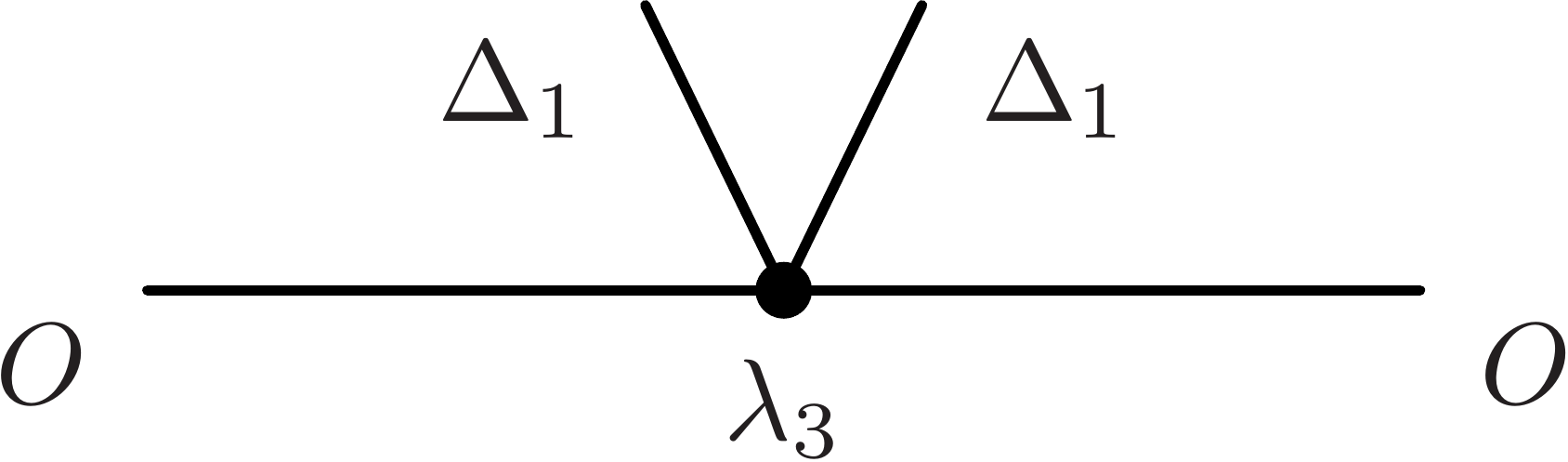}& \includegraphics[width=4.5cm]{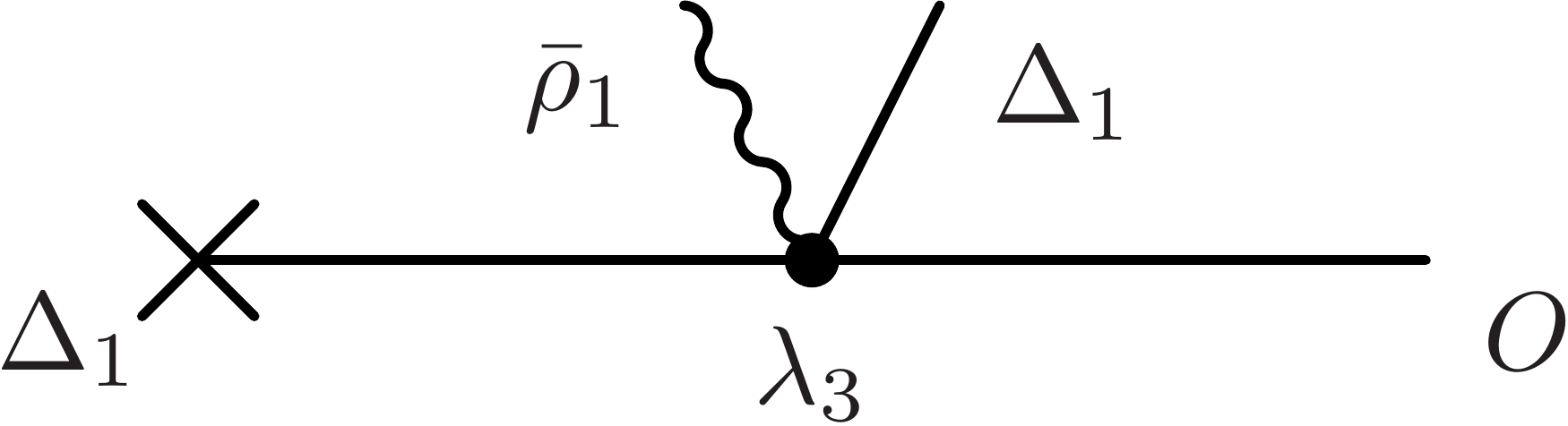} \\
\includegraphics[width=4.5cm]{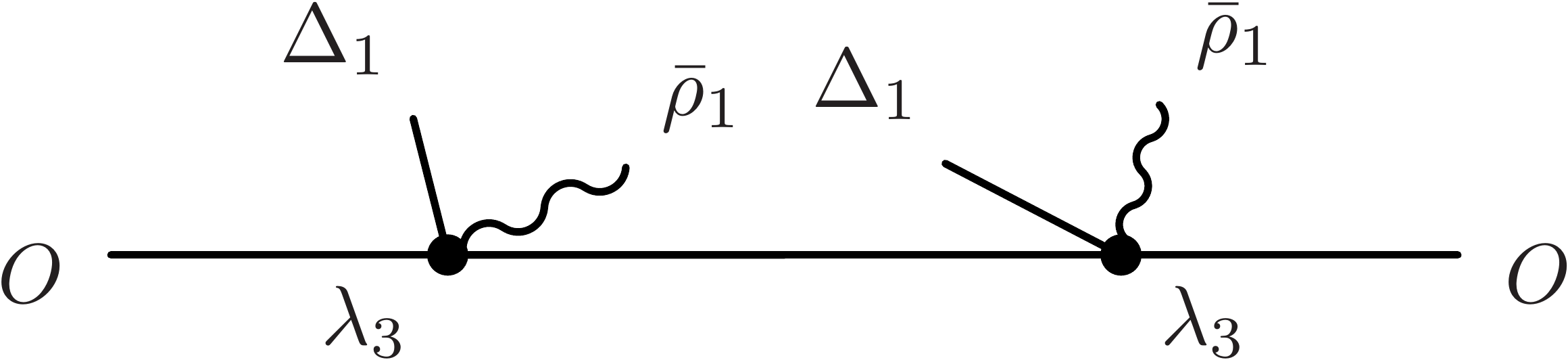} & \includegraphics[width=4.5cm]{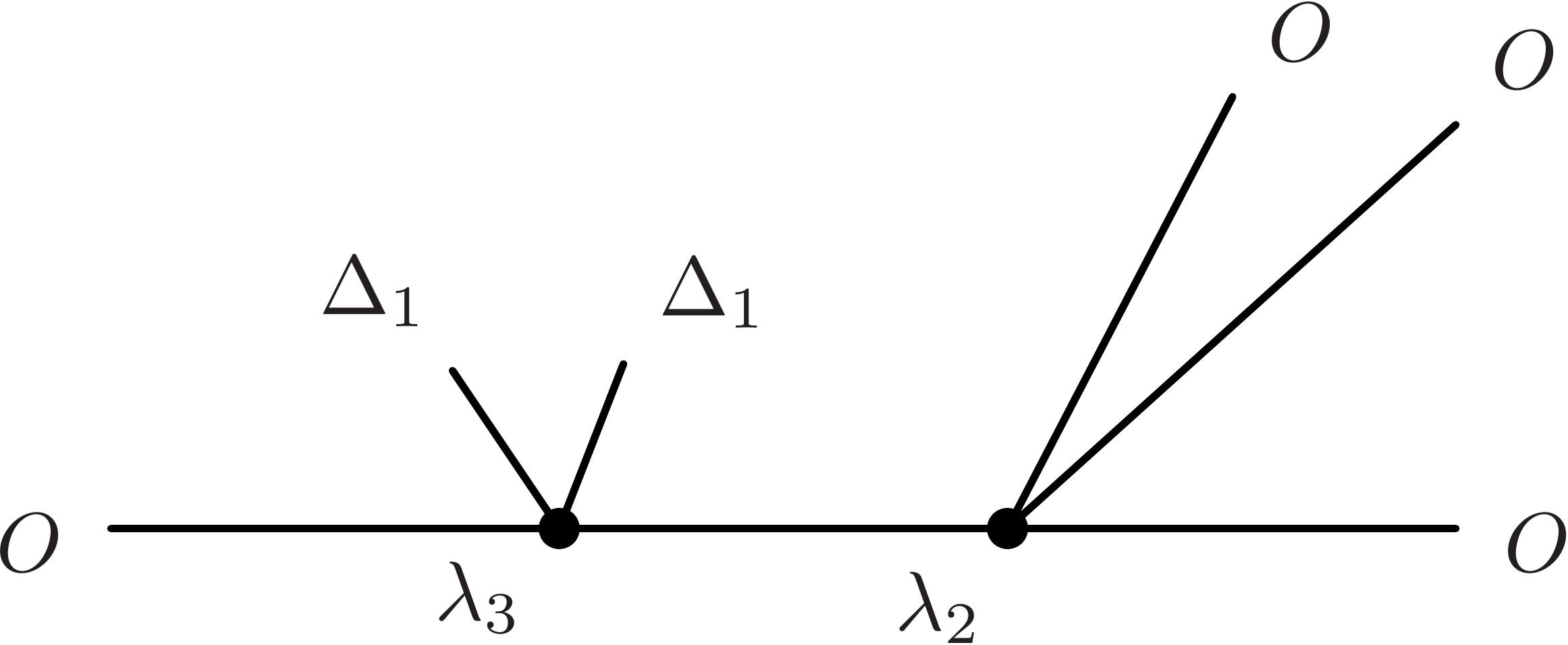}& \includegraphics[width=4.5cm]{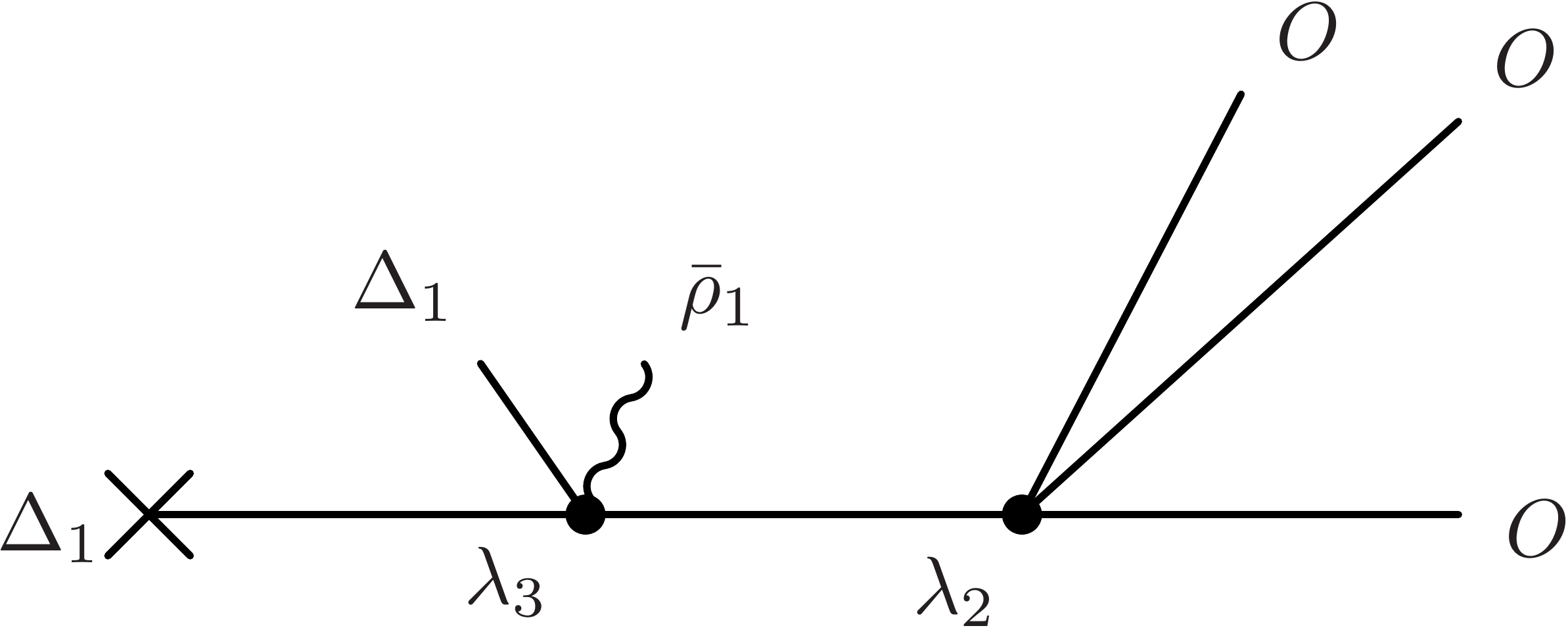} \\
 \includegraphics[width=4.5cm]{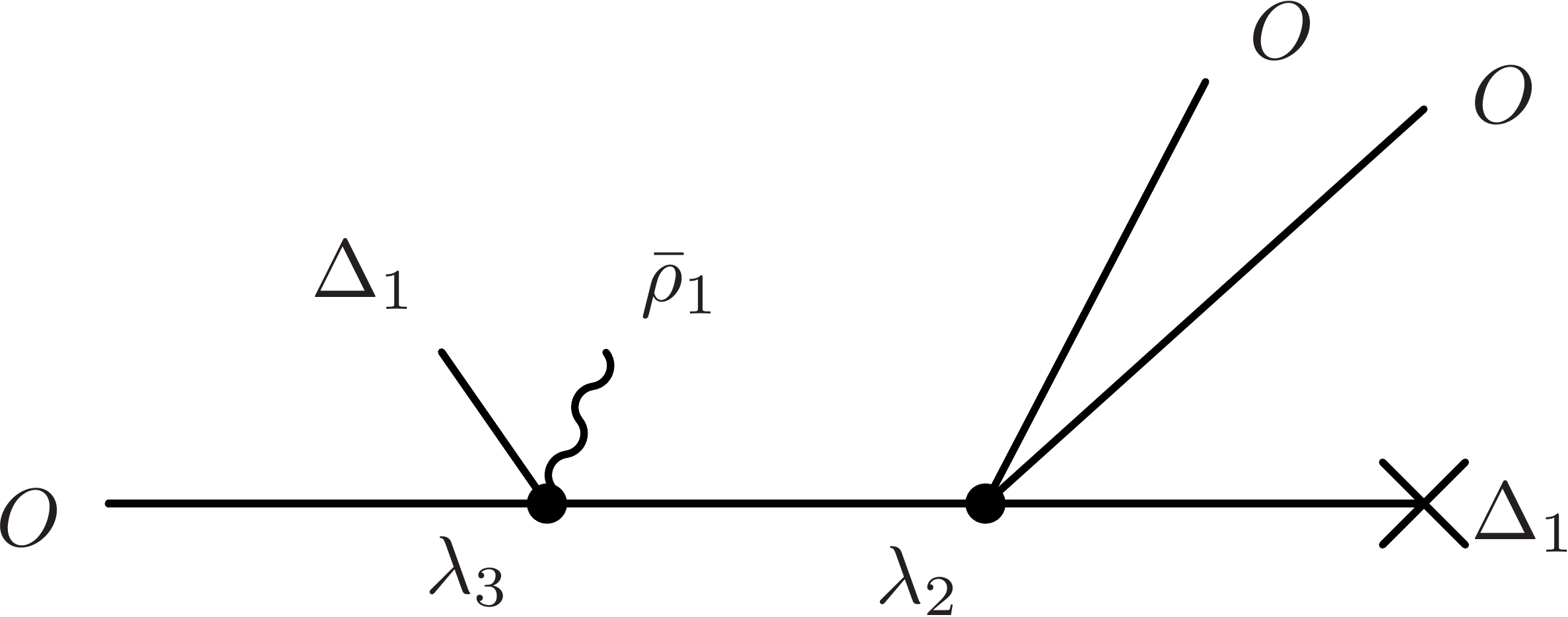}& \includegraphics[width=4.5cm]{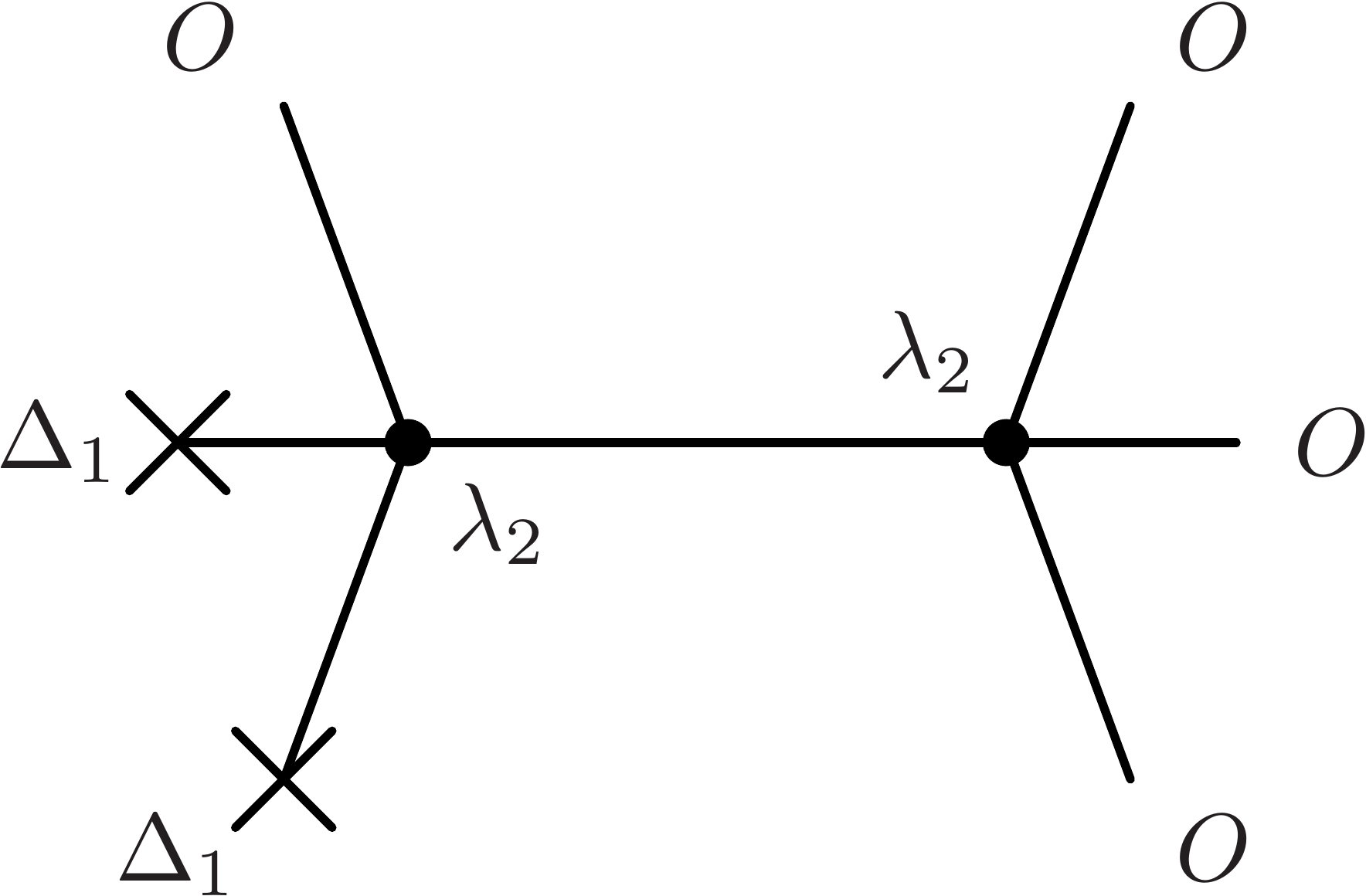}& \includegraphics[width=4.5cm]{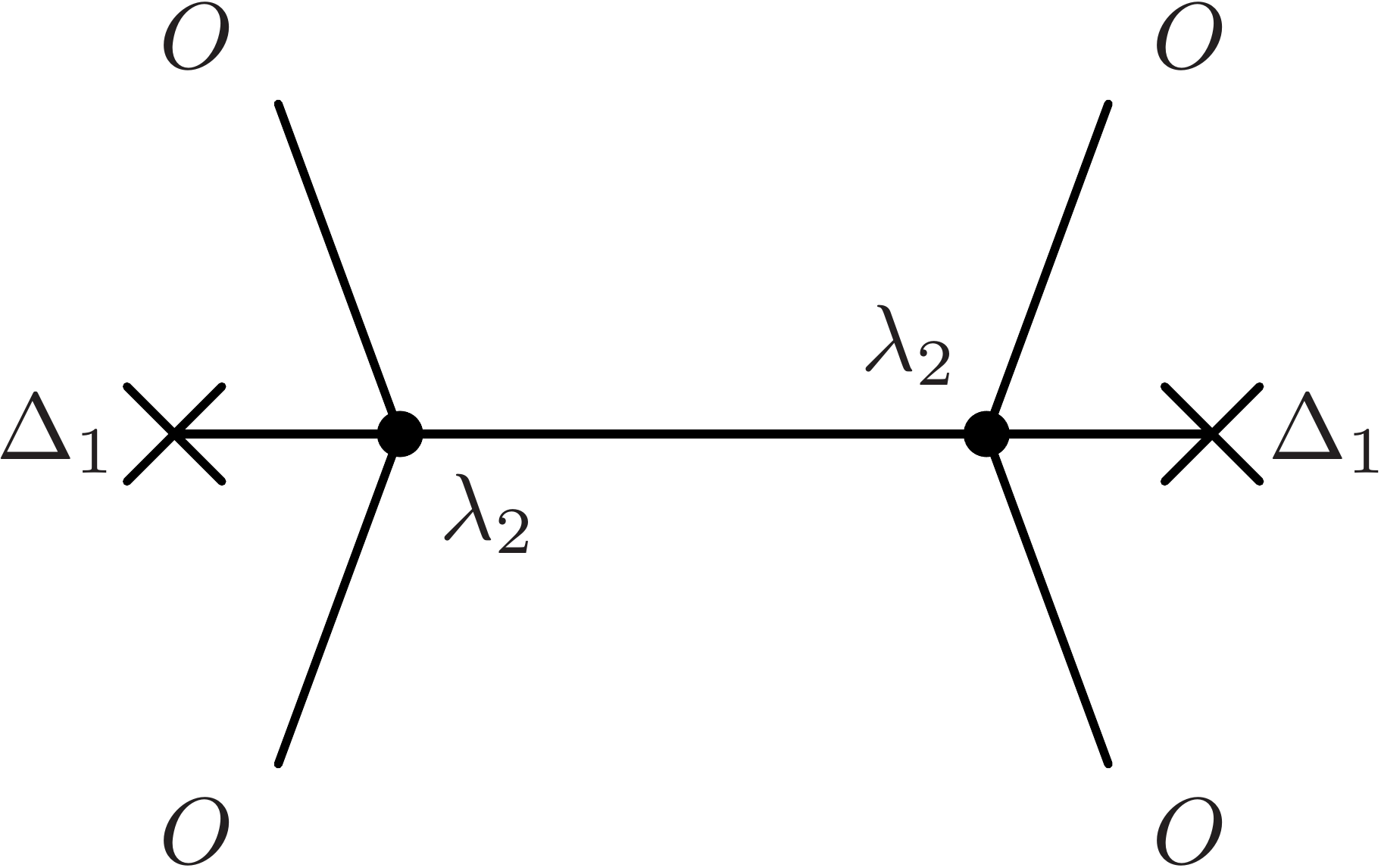} \\
\end{tabular}
\end{center}
\caption{From the figure of the upper left to that of the lower right, each Feynman diagram corresponds to each line of 
Eq.(\ref{eq:lint}) from the top line to the bottom. The cross mark denotes an insertion of the mass mixing term $m_{12}^2$. All the propagators are heavy scalars.} 
\label{fig:fig3}
\end{figure}
In Fig.\ref{fig:fig3}, we have shown the diagrams for Eq.(\ref{eq:lint}).
Below we calculate the contribution from $ \mathcal{L}_{\tmop{int}} (\Delta_1) $ in Eq.(\ref{eq:lint}).
In the last term of Eq.($\ref{eq: iGamma}$), the expression inside the logarithmic function is defined as  follows,
\begin{align}
	F [\bar{\rho}_1, n] 
		& \equiv \frac{\int d \Delta_1 e^{i \left\{ -\frac{1}{2} \int d^d x \int d^d y \Delta_1 (x) (D^{- 1}_{11})' (\bar{\rho}_1) 
			\Delta_1 (y) \right\}} e^{\left( \mathfrak{L}_{\tmop{int}} (\Delta_1) 
			- i \int d^d x \Delta_1 (x) \left( \frac{\delta \tilde{\Gamma}_{\tmop{eff}} [\bar{\rho}_1, n]}{\delta \bar{\rho}_1 (x)}\right)
			\right)}}{\int d \Delta_1 e^{i \left\{ \frac{1}{2} \int d^d x \int d^d y \Delta_1 (x) (D^{- 1}_{11})' (\bar{\rho}_1) 
			\Delta_1 (y) \right\}}}.
\end{align}
Note that the effect of $\int d^d x \Delta_1 (x) \left( \frac{\delta \tilde{\Gamma}_{\tmop{eff}} [\bar{\rho}_1, n]}{\delta \bar{\rho}_1 (x)}
\right)$ removes the contribution from one particle reducible graphs. 
We expand $F [\bar{\rho}_1, n]$ with respect to the series of the coupling constants.
Up to the second order, it is given by, 
\bea
F [\bar{\rho}_1, n] 
		& \approx & 1 +\frac{i}{4} \lambda_3 \mu^{2 \eta} \int d^d x \left(\prod_{i=1}^2 \int d^d x_i D^{(0)}_{F 22} (x - x_i) O (x_i) \right) 
			D'_{F 11} (x, x)  \notag \\
		& + & i \lambda_3 m_{12}^2 \mu^{2 \eta} \int d^d x \bar{\rho}_1 (x) 
			 \left(\prod_{i=1}^2 \int d^d x_i D^{(0)}_{F 22} (x - x_i) \right)  O (x_2) D'_{F 11} (x, x_1)  \notag \\
		& + & i \lambda_3 m_{12}^2 \mu^{2 \eta} \int d^d x \bar{\rho}_1 (x) \int d^d x_1 \int d^d x_2 D^{(1)}_{F 22} (x, x_1) 
			D^{(0)}_{F 22} (x - x_2) O (x_2) D'_{F 11} (x, x_1)  \notag \\
		& + & i \lambda_3 m_{12}^2 \mu^{2 \eta} \int d^d x \bar{\rho}_1 (x) \int  d^d x_1 \int d^d x_2 D^{(0)}_{F 22} (x - x_1) 
			D^{(1)}_{F 22} (x, x_2) O (x_2) D'_{F 11} (x, x_1)  \notag \\
		& + & \frac{\lambda_3^2\mu^{4 \eta}}{2}  \int d^d x \int d^d y \bar{\rho}_1 (x) \bar{\rho}_1 (y) D^{(0)}_{F 22} (x - y) \notag \\  
		& \times & \int d^d x_1 \int d^d y_1D^{(0)}_{F 22} (x - x_1) D^{(0)}_{F 22} (y - y_1) O (x_1) O (y_1) D'_{F 11} (x , y) .
			\label{F1stv1}
\eea
To derive Eq.(\ref{F1stv1}), we approximate the exact expression  by expanding the result with the small parameter   $\epsilon=\frac{m^2_{12}}{m^2_{2}} \ll 1$.
About the quadratic terms for the background field $\bar{\rho}_1$ ,  we keep the
the terms  suppressed up to the order of  $m^2_{12} \epsilon
\bar{\rho}_1^2 $.  About the quartic terms, we keep those suppressed by $\epsilon^2
 \bar{\rho}_1^4$. 
In Eq.(\ref{F1stv1}) , $D_{F22}^{(n)}(x,y), (n=0,1,2)$ are obtained by solving the following integral equation with iteration.
\begin{align} 
	D_{F22} (x, y) 
		& =  D_{F22}^{(0)} (x - y) - i \int d^d z D_{F22}^{(0)} (x - z)  \frac{\lambda_3}{2} \bar{\rho}_1^2 (z)  D_{F22} (z, y), \label{Df22expand0}
\end{align}
where the leading contribution $D_{F22}^{(0)} (x-y) $ is  given as follows,
\begin{align}
	D_{F22}^{(0)} (x-y) &= \int \frac{d^d k}{(2 \pi)^d i} \frac{1}{m_2^2 - k^2} e^{- ik  (x - y)}.
\end{align}
One also finds the first order correction and the second order correction. The same approximation is also used when we expand $D_{F22}$ in Eq.(\ref{W2tree1p}),
\begin{align}
	D_{F22}^{(1)} (x, y) 
		& =  - i \int d^d z D_{F22}^{(0)} (x - z) \frac{\lambda_3}{2} \bar{\rho}_1^2 (z) 
 D^{(0)}_{F22} (z - y), \\
	D_{F22}^{(2)} (x, z) 
		& =  - \int d^d z \int d^d \omega D_{F22}^{(0)} (x - z) \frac{\lambda_3}{2} \bar{\rho}_1^2 (z) D_{F22}^{(0)} (z - \omega) \frac{\lambda_3}{2} \bar{\rho}_1^2 (\omega) D^{(0)}_{F22} (\omega - y).
\end{align}

The propagator $D'_{F11} (x,y)$ is expanded with respect to $m_{12}$. When $\bar{\rho}_1$
is independent of the space time,  $D'_{F11} (x,y)$
 up to the fourth power of $m_{12}$ is given as follows.
\begin{eqnarray}
	D'_{F 11} (x-y) 	& = & \int \frac{d^d k}{(2\pi)^d} e^{-i k\cdot(x-y)} 
 \frac{- i}{m_1^2-k^2 + 3 \lambda_1 \bar{\rho}_1^2}  \nn \\ 
			&+& 
\int \frac{d^d k}{(2\pi)^d} e^{-i k\cdot(x-y)} 
\frac{- i m_{12}^4}{(m_1^2-k^2 + 3 \lambda_1 \bar{\rho}_1^2)^2 
			\left( m_2^2-k^2 + \frac{\lambda_3}{2} \bar{\rho}_1^2 \right)}. \label{Df11def}
\end{eqnarray}
Hereafter, we consider the case for the space time independent field
$\bar{\rho}_1$ and the fermion bilinear of $\bar{n}n$ for the neutrino.  
We study the effective potential for the scalar and the bilinear.
In order to study the scalar loop effect on the neutrinos Yukawa coupling,
 it is sufficient to consider the space time independent mode of the bilinear $\bar{n}n$.  
We summarize the tree level action and one loop contribution  for  the constant 
background 
fields.
\begin{eqnarray}
  \Gamma_{\tmop{eff}} [\bar{\rho}_1, n] & = & S [\bar{\rho}_1, 0, 0]+\tilde{\Gamma}_{\tmop{eff}}
  [\bar{\rho}_1, n] , \label{eq:bc} \\
  S [\bar{\rho}_1, 0, 0] & = & \int d^d x \left[ - \frac{Z_{m11}}{2} m_1^2
  \bar{\rho}_1^2 - \frac{\lambda_{01} {Z_1}^2 }{4}  \bar{\rho}_1^4 \right] \nn \\
  &+& \int d^dx \left[ \mu^{-2 \eta} \left(\sum_{i=1}^{2}Z_{h i} h_i {m_i}^4+ 
 2 Z_{h3} h_3 m^2_{1} m^2_{2} + Z_{h12} h_{12} m_{12}^4 \right) \right],
 \label{eq:S} \\ 
  \tilde{\Gamma}_{\tmop{eff}} [\bar{\rho}_1, n] & = &
  \tilde{\Gamma}_{\tmop{eff}} [\bar{\rho}_1, n]^{\tmop{tree}} +
  \tilde{\Gamma}_{\tmop{eff}} [\bar{\rho}_1, n]^{1 \tmop{loop}}_{\Delta_2} + \tilde{\Gamma}_{\tmop{eff}} [\bar{\rho}_1, n]^{1
  \tmop{loop}}_{\Delta_1} + \tilde{\Gamma}_{\tmop{eff}}[\bar{\rho}_1, n]^{\tmop{TrLn}}, \\
 \tilde{\Gamma}_{\tmop{eff}} [\bar{\rho}_1, n]^{\tmop{tree}}&=& 
 \int d^d x \left[ \frac{m_{012}^2}{2} \epsilon_0 \bar{\rho}_{01}^2 + y_0
  \epsilon_0 \bar{n_0} n_0 \bar{\rho}_{01} - \frac{\lambda_{03}}{4} \epsilon_0^2  
\bar{\rho}_{01}^4 + \frac{1}{2 m_2^2}  \left(
  \frac{\lambda_3 \mu^{2 \eta}}{2} \epsilon \bar{\rho}_1^3 - y \mu^{\eta} 
  \bar{n} n \right)^2 \right]    \label{eq:Gammatree}, 
 \\
  \tilde{\Gamma}_{\tmop{eff}} [\bar{\rho}_1, n]^{1 \tmop{loop}}_{\Delta_2}
  & = & \int d^d x \left[ 
 \frac{3 \lambda_2  \mu^{2 \eta}}{16 \pi^2}  ( \Gamma (\eta) + 1 +
  \log 4 \pi - \log m_2^2) \epsilon   
(y \mu^\eta \bar{n} n \bar{\rho}_1+\frac{m_{12}^2}{2}\bar{\rho}_1^2)
 \right. \nn \\
  &  &+ \left. \frac{3 \lambda_2 \lambda_3 \mu^{4 \eta}}{64 \pi^2}  (- \Gamma
  (\eta) - 2 - \log 4 \pi + \log m_2^2) \epsilon^2  \bar{\rho}_1^4 \right] , \label{eq:1loop1}
 \\
	\tilde{\Gamma}_{\tmop{eff}} [\bar{\rho}_1, n]^{1 \tmop{loop}}_{\Delta_1}
		& = & \int d^d x \left[ - \frac{\lambda_3 \mu^{2 \eta}}{16 \pi^2}  (\Gamma (\eta) + \log 4 \pi + 1 - \log m_2^2) 
			\epsilon (y \mu^\eta \bar{n} n \bar{\rho}_1+m_{12}^2 \bar{\rho}_1^2)  \right.\nn \\
		& + & \frac{\lambda_3 \mu^{2 \eta}}{16 \pi^2}  \left\{  \frac{3 \lambda_1 \mu^{2 \eta}}{4}  
			\left(  \Gamma (\eta) + 1 + \log 4 \pi - \log (m_1^2 + 3 \lambda_1 \mu^{2 \eta}  \bar{\rho}_1^2) 
			- 4 \log \left( \frac{m_1^2 + 3 \lambda_1 \mu^{2 \eta}  \bar{\rho}_1^2}{m_2^2} \right) \right) \right.\nn \\
		& + & \left.  \left. \lambda_3 \mu^{2 \eta}  \left( \Gamma (\eta) + \log 4 \pi + \frac{3}{2} - \log m_2^2 \right) \right\} 
			\epsilon^2  \bar{\rho}_1^4 \right] ,\label{eq:1loop2}\\
	\tilde{\Gamma}_{\tmop{eff}} [\bar{\rho}_1, n]^{\tmop{TrLn}}
		& = & \int d^d x \left[ \frac{m_{12}^4}{32 \pi^2}  (\Gamma [\eta] + \log 4 \pi + 1 - \log m_2^2) 
			+\frac{\epsilon^2 m_1^2m_2^2}{32\pi^2} \log \frac{m_1^2 + 3 \lambda_1\mu^{2\eta} \bar{\rho}_1^2}{m_2^2} \right.\nn \\
		& + & \frac{m_1^4}{64 \pi^2}  
			\left( \Gamma [\eta] + \log 4\pi+\frac{3}{2} - \log (m_1^2 + 3 \lambda_1  \bar{\rho}_1^2 \mu^{2 \eta}) \right) 
			+ \frac{m_2^4}{64 \pi^2}  \left( \Gamma [\eta] +\log 4\pi+ \frac{3}{2} - \log m_2^2 \right)\nn \\
		& + & \frac{m_{12}^2}{2}  
			\left( \frac{3 \lambda_1 \mu^{2 \eta}}{16 \pi^2} \log \frac{m_1^2 + 3 \lambda_1 \mu^{2 \eta}  \bar{\rho}_1^2}{m_2^2} 
			-\frac{\lambda_3 \mu^{2 \eta}}{32 \pi^2} \right) \epsilon \bar{\rho}_1^2 \nn \\
		& + & \left\{ \frac{3 \lambda_1 \mu^{2 \eta} m_1^2}{32 \pi^2}  \left( \Gamma [\eta] + \frac{3}{2} + \log 4 \pi 
			- \log (m_1^2 + 3 \lambda_1  \bar{\rho}_1^2 \mu^{2 \eta}) \right)\right. \nn \\
		& + & \left. \frac{\lambda_3 \mu^{2 \eta} m_2^2}{64 \pi^2}  \left( \Gamma [\eta] +1+ \log 4 \pi 
			- \log m_2^2 \right) \right\}  \bar{\rho}_1^2\nn \\
		& + & \left\{ \frac{\lambda_3^2 \mu^{2\eta}}{256 \pi^2} \left( \Gamma [\eta] +\log 4\pi - \log m_2^2  \right) \right. \nn \\
		& + & \left.  \frac{9 \lambda_1^2\mu^{4\eta}}{64 \pi^2} \left( \Gamma[\eta] + \frac{3}{2} +\log 4\pi
			- \log (m_1^2 + 3 \lambda_1 \bar{\rho}_1^2 \mu^{2 \eta})\right) \right\}  \bar{\rho}_1^4 \notag \\
		& +& \left.   \left\{ \frac{9\lambda_1^2\mu^{2\eta}}{32 \pi^2}\log \frac{m_1^2 + 3 \lambda_1\mu^{2\eta} \bar{\rho}_1^2}{m_2^2} -
			\frac{3\lambda_1\lambda_3 \mu^{4\eta} }{64\pi^2}
			\left(1+ \log \frac{m_1^2 + 3 \lambda_1\mu^{2\eta} \bar{\rho}_1^2}{m_2^2}   \right)
			+\frac{\lambda_3^2 \mu^{4\eta}}{256 \pi^2}\right\} \epsilon^2 \bar{\rho}_1^4  \right],
		\nn \\ \label{eq:TrLn}
\end{eqnarray}
where  $\epsilon_0=\frac{m_{012}^2}{m_{02}^2} $.  $\Gamma[\eta]$ is a gamma function which represents divergence in the dimensional regularization.
\section{The counter terms and the effective potential}
One loop contribution in Eqs.(\ref{eq:1loop1}-\ref{eq:TrLn}) includes the divergent terms. We show that the counter terms of  $S[\bar{\rho_1},0,0]$ in Eq.(\ref{eq:S}) and $\tilde{\Gamma}_{\tmop{eff}}^{\tmop{tree}}$ in Eq.(\ref{eq:Gammatree}) are  determined so that the divergences are canceled. 
We replace the bare mass terms and the bare coupling constants with the renormalized ones using 
the relations from
Eqs.(\ref{rho0i}-\ref{eq:cyukawa}). We also use the 
fact that there is no wave function renormalization for scalars from their one-loop 
diagrams.
Since we do not take into account of the fermion loop contribution, one can set $Z_i=1 (i=1,2)$. 
Since $m_1^2 m_2^2$ term does not include divergence, one can set $Z_{h_3}=1$.
The other counter terms
are generated by splitting the Z factors as , 
\begin{eqnarray}
  Z_{12} & = & 1 + (Z_{12} - 1), \\
  Z_{\lambda_{i i}} & = & 1 + (Z_{\lambda_{i i}} - 1) (i = 1, 2, 3) ,\\
  Z_{m i i} & = & 1 + (Z_{m i i} - 1) (i = 1, 2), \\
   Z_{h_i} & = & 1 + (Z_{h_i} - 1) (i = 1, 2), \quad  Z_{h_{12}} = 1 + (Z_{h_{12}} - 1).
\end{eqnarray}
As the result, the tree part of the effective action and  the counter terms are obtained as follows.
\begin{eqnarray}
	S_{\tmop{tree}}[\bar{\rho}_1, n]  &=&  S [\bar{\rho}_1, 0, 0] +  \tilde{\Gamma}_{\tmop{eff}} [\bar{\rho}_1, n]^{\tmop{tree}} 
		= S_{\tmop{tree}}^{\tmop{4 dim}}[\bar{\rho}_1, n]+S_{\tmop{tree}}^{\tmop{6 dim}}[\bar{\rho}_1, n]+S_{\tmop{C}}[\bar{\rho}_1,n],\\
 S_{\tmop{tree}}^{\tmop{4 dim}}[\bar{\rho}_1, n]
		&=&   \int d^d x \left[ - \frac{1}{2} m_1^2 \bar{\rho}_1^2 - \frac{\lambda_1}{4}  \bar{\rho}_1^4
		+\frac{m_{12}^2}{2} \epsilon \bar{\rho}_1^2 + y \mu^{\eta} \epsilon \bar{n} n \bar{\rho}_1 
		- \frac{\lambda_3 \mu^{2 \eta}}{4} \epsilon^2  \bar{\rho}_1^4  \right] \nn \\
		& + & \int d^d x \mu^{-2 \eta} \left[\sum_{i=1}^{2} h_i {m_i}^4+ 
 2 h_3 m^2_{1} m^2_{2} + h_{12}m_{12}^4 \right]
 \\ 
	S_{\tmop{tree}}^{\tmop{6 dim}}[\bar{\rho}_1, n]
		 &=&   \int d^d x \left[ \frac{1}{2 m_2^2}  \left(\frac{\lambda_3 \mu^{2 \eta}}{2} \epsilon \bar{\rho}_1^3 
		- y \mu^{\eta} \bar{n} n \right)^2 \right], \\
 	S_{\tmop{C}}[\bar{\rho}_1, n]&=&
  \int d^d x \left[  \left( - \frac{(Z_{m 11} - 1) m_1^2 + Z_{m 12} m_2^2}{2} \bar{\rho}_1^2 
 - \frac{\{ (Z_{\lambda_{11}} - 1) \lambda_1 + Z_{\lambda_{12}} \lambda_2 + Z_{\lambda_{13}} \lambda_3 \} \mu^{2 \eta}}{4} 
		\bar{\rho}_1^4 \right)  \right.  \nn \\
		&+& \frac{1}{2} m_{12}^2 \left( 2 (Z_{12} - 1) - \left\{ (Z_{m 22} - 1) 
		 \right\} \right) \epsilon \bar{\rho}_1^2 
		 + \left( (Z_{12} - 1) - \left\{ (Z_{m 22} - 1) 
		\right\} \right) y \mu^{\eta} \epsilon \bar{n} n \bar{\rho}_1 \nn \\
		&-& \frac{\mu^{2 \eta}}{4} \left[ Z_{\lambda_{31}} \lambda_1 + Z_{\lambda_{32}} \lambda_2 
		+ (Z_{\lambda_{33}} - 1) \lambda_3 + 2\lambda_3 \left\{ (Z_{12} - 1) - \left( (Z_{m 22} - 1) 
		 \right) \right\} \right] \epsilon^2 \bar{\rho}_1^4 \nn \\ 
		 & + & \left. \mu^{-2 \eta} \left(\sum_{i=1}^{2}(Z_{h i}-1) h_i {m_i}^4+ (Z_{h12}-1)h_{12}m_{12}^4 \right) \right]. 
\label{eq:Sc}
\end{eqnarray}
We keep the terms up to those suppressed as  $\epsilon \bar{\rho}_1^2$ , $\epsilon^2 \bar{\rho}_1^4$ and $\epsilon \bar{n} n \bar{\rho}_1$
and ignore the terms with the further suppression factor of $\frac{m_1^2}{m_2^2}$. Correspondingly, the counter terms  with the same suppression factor such as  $Z_{m21}\frac{m_1^2}{m_2^2}$ are also ignored in Eq.(\ref{eq:Sc}).  The $Z$ factors in Eq.(\ref{eq:Sc}) are determined in the full theory and the derivation is given 
 in appendix A.  By substituting the $Z$ factors in Eqs.(\ref{eq:Zm21}-\ref{eq:fullcounter}), all the divergences in Eqs.(\ref{eq:1loop1}-\ref{eq:TrLn}) are canceled.
We define the tree level part and the one loop part of the effective action respectively as,
\bea
S_{\tmop{eff}}^{\tmop{tree}}&=&S_{\tmop{tree}}^{\tmop{4 dim}}[\bar{\rho}_1, n]+ S_{\tmop{tree}}^{\tmop{6 dim}}[\bar{\rho}_1, n],  \label{eq: Sefftree}\\
S_{\tmop{eff}}^{\tmop{loop}}&=& \tilde{\Gamma}_{\tmop{eff}} [\bar{\rho}_1, n]^{1 \tmop{loop}}_{\Delta_2} + 
\tilde{\Gamma}_{\tmop{eff}} [\bar{\rho}_1, n]^{1\tmop{loop}}_{\Delta_1} + \tilde{\Gamma}_{\tmop{eff}}[\bar{\rho}_1, n]^{\tmop{TrLn}}
+S_{\tmop{C}}[\bar{\rho}_1, n], \label{eq:Seff1loop} \\
 S_{\tmop{eff}} & = & S_{\tmop{eff}}^{\tmop{tree}} +
 S_{\tmop{eff}}^{\tmop{loop}} \nn \\
	& = & \int d^4 x \Bigl[  
		m_1^4\left\{ h_1+ \frac{1}{64 \pi^2}  \left( \frac{3}{2} - \log \frac{m_1^2 + 3 \lambda_1  \bar{\rho}_1^2 }{\mu^2}  \right) \right\} 
		+m_2^4 \left\{  h_2  + \frac{1}{64 \pi^2}  \left( \frac{3}{2} - \log \frac{m_2^2}{\mu^2} \right) \right\} 
		\nn \\
	& + & m_{12}^4 \left\{ h_{12} + \frac{1}{32 \pi^{2}}  \left( 1 - \log \frac{m_2^2}{\mu^2} \right) \right\}
		 + 2 m_1^2 m_2^2 \left\{ h_3 +\frac{\epsilon^2}{32\pi^2}  \log  \frac{m_1^2 + 3 \lambda_1  \bar{\rho}_1^2 }{\mu^2}  \right\}   
		 \nn \\ 
	& - & \frac{1}{2} \left\{ m_1^2 \left( 1 - \frac{3 \lambda_1 }{16 \pi^2}  
		\left( \frac{3}{2} - \log \frac{m_1^2 + 3 \lambda_1  \bar{\rho}_1^2 }{\mu^2} \right) \right) 
		- \frac{\lambda_3 m_2^2}{32 \pi^2}  \left( 1 - \log \frac{m_2^2}{\mu^2} \right) \right\} \bar{\rho}_1^2 \nn \\
	& + & \frac{m_{12}^2}{2} \left\{ 1 + \frac{3 \lambda_1}{16 \pi^2} \log \frac{m_1^2 + 3 \lambda_1   \bar{\rho}_1^2}{m_2^2} 
		+\frac{3 \lambda_2}{16 \pi^2}  \left( 1 - \log \frac{m_2^2}{\mu^2} \right) 
		- \frac{\lambda_3 }{8 \pi^2}  \left( \frac{5}{4} - \log \frac{m_2^2}{\mu^2} \right) \right\} \epsilon \bar{\rho}_1^2 \nn \\
	& - & \frac{\lambda_1 }{4} \left\{ 1 - \frac{9 \lambda_1 }{16 \pi^2} 
		\left( \frac{3}{2} - \log  \frac{m_1^2 + 3 \lambda_1  \bar{\rho}_1^2 }{\mu^2}  \right) 
		+ \frac{1}{64 \pi^2} \frac{\lambda_3^2 }{\lambda_1} \log \frac{m_2^2}{\mu^2}  \right\} \bar{\rho}_1^4 \nn \\
	& - & \frac{\lambda_3 }{4}  \left\{ 1 + \frac{3 \lambda_2 }{16 \pi^2}  
		\left( 2 - \log \frac{m_2^2}{\mu^2} \right) \right. 
		- \frac{\lambda_3}{4 \pi^2}  \left( \frac{25}{16} - \log \frac{m_2^2}{\mu^2} \right)\nn \\
	& + & \left. \frac{3 \lambda_1}{16 \pi^2}  \left(  \log  \frac{m_1^2 + 3 \lambda_1   \bar{\rho}_1^2}{\mu^2} 
		+ 5 \log  \frac{m_1^2 + 3 \lambda_1   \bar{\rho}_1^2}{m_2^2} 
		 -6\frac{\lambda_1}{\lambda_3} \log  \frac{m_1^2 + 3 \lambda_1   \bar{\rho}_1^2}{m_2^2}\right) \right\}
		\epsilon^2  \bar{\rho}_1^4 \nn \\
	& + & \left. \left\{ 1 + 
		\frac{3 \lambda_2 - \lambda_3}{16 \pi^2}  \left( 1 - \log \frac{m_2^2}{\mu^2} \right) \right\} y \epsilon \bar{n} n \bar{\rho}_1
		+\frac{\lambda_3^2 }{8 m_2^2} \epsilon^2 \bar{\rho}_1^6 - \frac{\lambda_3 y }{2 m_2^2} \epsilon \bar{n} n \bar{\rho}_1^3 
		+ \frac{y^2  }{2 m_2^2} (\bar{n} n)^2 \right],
\end{eqnarray}
where the limit $d\to4$ is taken.
By substituting the vacuum expectation value $v_1$ for $\bar{\rho}_1$, one obtains the
effective potential as;
\bea
	V_{\mathrm{eff}}(v_1)&=&V_{\tmop{cosmo}} +\frac{m_{1\mathrm{eff}}^2}{2} v_1^2
		-\frac{m_{12 \mathrm{eff} }^2}{2} \epsilon v_1^2 +\frac{\lambda_{1\mathrm{eff}}}{4} v_1^4
		+\frac{\lambda_{3 \mathrm{eff}}}{4}  \epsilon^2 v_1^4 -y_{\mathrm{eff}} \epsilon \bar{n}n v_1 \nn \\
		&-&\frac{\lambda_3^2 }{8 m_2^2} \epsilon^2 v_1^6 
		+\frac{\lambda_3}{2m_2^2}  \epsilon (\bar{n}n) v_1^3 
		-\frac{y^2}{2m_2^2} (\bar{n}n)^2. \label{Veffective} 
\eea
where the cosmological constant, the effective masses and  couplings are defined as;
\bea
 V_{\tmop{cosmo}} & = & -h_{1\mathrm{eff}}{m_1}^4-h_{2 \mathrm{eff}} {m_2}^4-h_{12 \mathrm{eff}} {m_{12}}^4 -2 h_{3 \mathrm{eff}} {m_1}^2 {m_2}^2,
\label{Vcosmo}
\\
m_{1 \mathrm{eff}}^2&=& m_1^2 
\left( 1 - \frac{3 \lambda_1 }{16 \pi^2}  
		\left( \frac{3}{2} - \log \frac{m_1^2 + 3 \lambda_1  v_1^2 }{\mu^2} \right) \right) 
-  \frac{\lambda_3 m_2^2}{32 \pi^2}  \left( 1 - \log \frac{m_2^2}{\mu^2} \right), \label{m1eff} \\
m_{12\mathrm{eff}}^2&=& m_{12}^2
\left\{ 1 + \frac{3 \lambda_1}{16 \pi^2} \log
 \frac{m_1^2 + 3 \lambda_1   v_1^2}{m_2^2}  
		+\frac{3 \lambda_2}{16 \pi^2}  \left( 1 - \log \frac{m_2^2}{\mu^2} \right) 
		- \frac{\lambda_3 }{8 \pi^2}  \left( \frac{5}{4} - \log \frac{m_2^2}{\mu^2} \right) \right\}, \nn \\
\label{m12eff} \\
\frac{\lambda_{1\mathrm{eff}}}{4}&=&\frac{\lambda_1}{4}
\left\{ 1 - \frac{9 \lambda_1 }{16 \pi^2} 
		\left( \frac{3}{2} - \log  \frac{m_1^2 + 3 \lambda_1  v_1^2 }{\mu^2}  \right) 
		+ \frac{1}{64 \pi^2} \frac{\lambda_3^2 }{\lambda_1} \log \frac{m_2^2}{\mu^2}  \right\} , \label{l1eff} \\
\frac{\lambda_{3 \mathrm{eff}}}{4}&=&\frac{\lambda_3}{4} 
\left\{ 1 + \frac{3 \lambda_2 }{16 \pi^2}  
		\left( 2 - \log \frac{m_2^2}{\mu^2} \right) \right. 
		- \frac{\lambda_3}{4 \pi^2}  \left( \frac{25}{16} - \log \frac{m_2^2}{\mu^2} \right) \nn  \\
	& + & \left. \frac{3 \lambda_1}{16 \pi^2}  \left(  \log  \frac{m_1^2 + 3 \lambda_1   v_1^2}{\mu^2} 
		+ 5 \log  \frac{m_1^2 + 3 \lambda_1   v_1^2}{m_2^2} 
		 -6\frac{\lambda_1}{\lambda_3} \log  \frac{m_1^2 + 3 \lambda_1   v_1^2}{m_2^2}\right) \right\} , \label{l3eff} \\
h_{1\mathrm{eff}}&=& h_1+\frac{1}{64 \pi^2
}\left(\frac{3}{2}-\log \frac{m_1^2+3 \lambda_1 v_1^2 }{\mu^2}\right), \label{h1eff}
 \\
h_{2\mathrm{eff}}&=& h_2+\frac{1}{64 \pi^2} \left(\frac{3}{2}-\log\frac{m_2^2}{\mu^2} \right), \label{h2eff} \\
h_{3 \mathrm{eff}}&=& h_3 +\frac{\epsilon^2}{32\pi^2}  \log  \frac{m_1^2 + 3 \lambda_1  \bar{\rho}_1^2 }{\mu^2}, 
\\ 
h_{12 \mathrm{eff}}&=&h_{12}+\frac{1}{32\pi^2} \left(1-\log\frac{m_2^2}{\mu^2}  \right) ,
\label{h12eff} \\
y_{\mathrm{eff}}&=& y \left( 1+\frac{3\lambda_2-\lambda_3}{16 \pi^2}\left(1-\log\frac{m_2^2}{\mu^2} \right) \right). \label{yeff}
\eea
The renormalization scale $\mu$ independence  of the effective mass, coupling constant and cosmological constant is studied in the appendix A.
In the appendix, we have shown,
\bea
 &&  \mu \frac{d m_{1 \tmop{eff}}^2}{d \mu} =0, \quad   \mu \frac{d \lambda_{1 \tmop{eff}}}{d \mu}   = 0, \quad \mu \frac{dV_{\mathrm{cosmo}}}{d\mu}=0.
\eea
The other parameters are approximately scale independent. 
 Within the leading order of the expansion with respect to $\frac{m_1^2}{m_2^2}$,
they satisfy,
\bea
 && \mu \frac{d (m_{12 \tmop{eff}}^2 \epsilon)}{d \mu} =  m_{12}^2 \epsilon  O \left(
  \frac{m_1^2}{m_2^2} \right)\simeq 0,\\
 && 
  \mu \frac{d (\lambda_{3 \tmop{eff}} \epsilon^2)}{d \mu}  = \epsilon^2 O \left( \frac{m_1^2}{m_2^2} \right) \simeq 0, \\
  && \mu \frac{d (y_{\tmop{eff}} \epsilon)}{d \mu} = \epsilon y O
  \left( \frac{m_1^2}{m_2^2} \right) \simeq 0.
\end{eqnarray}
\section{RG improvement}
In this section, we discuss the RG improvement of the effective potential of Eqs.(\ref{Veffective}-\ref{yeff}). 
The RG improved effective potential for the models with two scalars is studied in \cite{Manohar:2020nzp, Okane:2019npj}.
The RG improved effective potential with multi-scale is also studied in \cite{Chataignier:2018aud}.

By setting the renormalization scale $\mu$  equal to the heavy 
scalar mass $m_2$, 
the obtained effective couplings and masses in Eqs.(\ref{m1eff}-\ref{h1eff}) include the large logarithmic correction 
which is proportional to $\log \frac{m_2^2}{m_1^2+3\lambda_1 v_1^2}$. In this section, we will resum this type of logarithmic corrections. 
Since their origin is  the loop correction of the light scalar whose virtual momentum ranges  from the heavy scalar mass $m_2$ down to the low energy, 
one can compute the correction by using the
effective low energy theory without the heavy scalar. 
We derive the low energy effective Lagrangian  by  integrating the tree level contribution of the heavy scalar in Eq.(\ref{W2tree1p}),
\bea
S&=&S[\bar{\rho}_1]+\bar{W}^{c \ \rm{tree}}_2(\bar{\rho}_1, n) \nn \\
 &=& \int d^dx  \left\{ \frac{1}{2} \partial_\mu \bar{\rho}_1  \partial^\mu \bar{\rho}_1 -\frac{m_1^2}{2} {\bar{\rho}_1}^2
-\frac{\lambda_1}{4}  {\bar{\rho}_1}^4+ 
h_1 {m_1}^4  +h_2 {m_2}^4 + 2 h_3 {m_1}^2{m_2}^2 +  {h_{12}}m_{12}^4
\right\}  \nn \\
&+& \frac{i}{2} \int d^d x d^d y O(x) D_{F22}(x,y) O(y),
\eea
where $D_{F22}(x,y)$ has the following form of the low energy expansion,
\bea
D_{F22}(x,y)=\frac{1}{i}\left(\frac{1}{m_2^2}-
\frac{\Box_x+\frac{\lambda_3}{2} \bar{\rho_1}^2}{m_2^4}+
\frac{(\Box_x+\frac{\lambda_3}{2} \bar{\rho_1}^2)^2 }{m_2^6}
 \right) \delta^d(x-y).
\eea
If we keep the terms up to  the dimension six ($d=6$) operators, the effective action is given as,
\bea
S&=&\int d^dx  \left\{ \frac{1}{2} \partial_\mu \bar{\rho}_1  \partial^\mu \bar{\rho}_1 -\frac{m_1^2}{2} {\bar{\rho}_1}^2
-\frac{\lambda_1}{4}  {\bar{\rho}_1}^4
+ \frac{m_{12}^4}{2 m_2^2}  {\bar{\rho}_1}^2+y \frac{m^2_{12}}{m_2^2} \overline{n}{n} \bar{\rho}_1  \right.
\nn \\
&+& 
h_1 {m_1}^4  +h_2 {m_2}^4 + 2 h_3 {m_1}^2{m_2}^2 +  {h_{12}}m_{12}^4 \nn \\
&+&\left. \frac{y^2}{2 m_2^2} (\overline{n}{n})^2 +  \frac{m_{12}^4}{2 m_2^4}  \partial_\mu \bar{\rho}_1  \partial^\mu \bar{\rho}_1 - \frac{\lambda_3}{4}\frac{m_{12}^4}{m_2^4}
\bar{\rho}_1^4
-y \frac{m^2_{12}}{m_2^2} \frac{\Box \bar{\rho}_1+\frac{\lambda_3 \bar{\rho}^3_1}{2}} {m_2^2}(\overline{n} n) 
+\frac{m_{12}^4 (\Box \bar{\rho}_1+\frac{\lambda_3 \bar{\rho}_1^3}{2})^2}{ 2 m_2^6}
 \right\}.\nn \\
\eea
One rewrites the action with $\epsilon=\frac{m^2_{12}}{m_2^2}$ and the rescaled field
$\rho_1^\prime =  \sqrt{1 + \epsilon^2} \rho_1$.
\bea
S&=&\int d^d x \left\{ \frac{1}{2} \partial_{\mu} \overline{\rho}^\prime_1
   \partial_{}^{\mu} \overline{\rho}^\prime_1 - \frac{m_1^2 - \epsilon m_{12}^2}{2}
   {\overline{\rho}^\prime_1}^2 - \frac{\lambda_1 + \epsilon^2 (\lambda_3 - 2
   \lambda_1)}{4} {\overline{\rho}^\prime_1}^4 + \epsilon y \bar{n} n
   {\overline{\rho}^\prime_1}  \right. \nn \\
&-& \left. \epsilon y \bar{n} n \frac{\Box
   \overline{\rho}^\prime_1}{m_2^2} + \frac{1}{2 m_2^2} \left( \frac{\epsilon
   \lambda_3 {\overline{\rho}^\prime_1}^3}{2} - y \bar{n} n \right)^2 +
   \frac{(\epsilon \Box {\overline{\rho}^\prime_1})^2}{2 m_2^2} + \frac{(\epsilon^2
   \Box {\overline{\rho}^\prime_1})}{2 m_2^2} \lambda_3 {\overline{\rho}^\prime_1}^3  \right. \nn \\
&+&\left.  h_1 {m_1}^4  +h_2 {m_2}^4 + 2 h_3 {m_1}^2{m_2}^2 +  {h_{12}}m_{12}^4   
\right\} .
\label{eqlowL}
\eea
Below we derive the effective potential including one loop effects of light scalar and improve it with RG equation.
As for $d=6$ operators, we estimate  the one loop contribution to the renormalizable terms in the effective potential. As the result, the contribution turned out to be suppressed by the higher powers of $\epsilon$ and $\frac{m_1^2}{m_2^2}$. Then  
ignoring this contribution, we will obtain the effective potential  within the following accuracy:
(1) For higher dimensional  ($d=6$) terms, we calculate the contribution within the tree level approximation.
(2) For the renormalizable part of the effective potential, 
 the one-loop contribution is included. 
The tree level effective potential  is given by substituting the constant vacuum expectation value  $(1+\epsilon^2)v_1^2$ to $\bar{\rho_1}^{\prime 2}$  in  Eq.(\ref{eqlowL}).
\bea
V_{\tmop{eff}}^{\tmop{tree}}&=&\frac{m_1^2 - \epsilon m_{12}^2}{2} {v_1}^2 + 
\frac{\lambda_1 + \epsilon^2 \lambda_3 }{4}{v_1}^4 -\epsilon y \bar{n} n
   {v_1}  \nn \\
&-& h_1 {m_1}^4  -h_2 {m_2}^4 - 2 h_3 {m_1}^2{m_2}^2 - {h_{12}}m_{12}^4   \nn \\
&-& \frac{1}{2 m_2^2} \left( \frac{\epsilon
   \lambda_3 {v_1}^3}{2} - y \bar{n} n \right)^2 .
\eea
The total contribution including the one-loop correction and its counter terms is given as;
\bea
V^{\mathrm{Low}}_{\tmop{eff}}&=&V_{\tmop{eff}}^{\tmop{tree}}+V_{\tmop{eff}}^{\tmop{1 loop}}
 + V^c_{\tmop{eff}} \nn \\
&=& 
\frac{m_1^2}{2} \left\{ 1 - \frac{3 \lambda_1}{16 \pi^2} \left(
  \frac{3}{2} - \log \left( \frac{m_1^2 + 3 \lambda_1 v_1^2}{\mu^2} \right)
  \right) \right\} v_1^2 \nn \\
&& - \frac{m_{12}^2}{2} \left\{ 1 - \frac{3
  \lambda_1}{16 \pi^2} \left( 1 - \log \left( \frac{m_1^2 + 3 \lambda_1
  v_1^2}{\mu^2} \right) \right) \right\} \epsilon v_1^2  - \epsilon y \bar{n} n v_1 - \frac{1}{2 m_2^2} \left( \frac{\lambda_3 \epsilon}{2} v_1^3 - y
  \bar{n} n \right)^2
   \nn \\
&& + \frac{\lambda_1}{4} \left\{ 1 - \frac{9 \lambda_1}{16 \pi^2} \left(
  \frac{3}{2} 
- \log \left( \frac{m_1^2 + 3 \lambda_1 v_1^2}{\mu^2} \right)
  \right) \right\} v_1^4 \nn \\
&& + \frac{\lambda_3}{4} \left\{ 1 + \frac{9}{8 \pi^2}
  \frac{\lambda_1 (\lambda_1 - \lambda_3)}{\lambda_3} + \frac{9}{8 \pi^2}
  \frac{\lambda_1 (\lambda_3 - \lambda_1)}{\lambda_3} \log \left( \frac{m_1^2
  + 3 \lambda_1 v_1^2}{\mu^2} \right) \right\} \epsilon^2 v_1^4 \nn \\
  &  & - m_1^4 \left\{ h_1 + \frac{1}{64 \pi^2} \left( \frac{3}{2} - \log
  \left( \frac{m_1^2 + 3 \lambda_1 v_1^2}{\mu^2} \right) \right) \right\}  \nn \\
&& +2
  \epsilon^2 m_1^2 m_{2}^2 \left\{ -\frac{h_3}{\epsilon^2} + \frac{1}{64 \pi^2} \left( 1 - \log
  \left( \frac{m_1^2 + 3 \lambda_1 v_1^2}{\mu^2} \right) \right) \right\} \nn \\
&& - \epsilon^2 m_{12}^4 \left\{ \frac{h_{12}}{\epsilon^2} - \frac{1}{64 \pi^2} \log \left(
  \frac{m_1^2 + 3 \lambda_1 v_1^2}{\mu^2} \right)  \right\} -h_2 {m_2}^4 .
\label{eq:veff}
\eea
The derivation of Eq.(\ref{eq:veff}) is given in appendix B. $V_{\tmop{eff}}^{\tmop{1 loop}}$ and $V^c_{\tmop{eff}}$ can be found in Eq.(\ref{eq:Veff1loop}) and in 
Eq.(\ref{eq:countertermforveff}), respectively.
To obtain the effective potential Eq.(\ref{eq:veff}) from Eq.(\ref{eq:veff0}),  we keep the terms such as $m_1^2
v_1^2 \epsilon^0$,  $m_{12}^2 v_1^2 \epsilon$ , $ v_1^4$ and $\epsilon^2 v_1^4$. 
 About the cosmological constant terms,
 we keep the terms of the forms $m_1^4 \epsilon^0$, $m_1^2
m^2_{12} \epsilon $ and $m_{12}^4 \epsilon^2 $. 
We drop the other terms with the extra suppression factors.

We compare the low energy effective potential in  Eq.(\ref{eq:veff}) to
the effective potential  Eq.(\ref{Veffective}). The latter  includes both heavy and light scalar loop effect . 
We set the renormalization scale  $\mu$ equal to $m_2$ 
in both effective potentials.
For $\mu = m_2$, the effective potential in Eq.(\ref{Veffective}) becomes,
\bea
	V_{\mathrm{eff}}& = & \frac{m_1^2}{2} \left[ 1 - \frac{3 \lambda_1}{16 \pi^2} \frac{3}{2} - \frac{\lambda_3}{32 \pi^2} \frac{m_2^2}{m_1^2} 
		+\frac{3 \lambda_1}{16 \pi^2} \log \left( \frac{m_1^2 + 3 \lambda_1 v_1^2}{m_2^2} \right) \right] v_1^2 \nn \\
	&  &	- \frac{m_{12}^2}{2} \left[1 - \frac{5 \lambda_3}{32 \pi^2} + \frac{3 \lambda_2}{16 \pi^2} 
		+\frac{3 \lambda_1}{16 \pi^2} \log \left( \frac{m_1^2 + 3 \lambda_1 v_1^2}{m_2^2} \right) \right] \epsilon v_1^2 \nn \\
	&  & + \frac{\lambda_1}{4} \left[ 1 - \frac{9 \lambda_1}{16 \pi^2} \frac{3}{2} 
		+\frac{9 \lambda_1}{16 \pi^2} \log \frac{m_1^2 + 3 \lambda_1 v_1^2}{m_2^2} \right] v_1^4 \nn \\
	&  &	+ \frac{\lambda_3}{4} 
		\left[ 1 + \frac{6 \lambda_2}{16 \pi^2} - \frac{25 \lambda_3}{64 \pi^2} 
		\frac{9 \lambda_1}{8 \pi^2}\left( 1 - \frac{\lambda_1}{\lambda_3} \right) 
		\log \left( \frac{m_1^2 + 3 \lambda_1 v_1^2}{m_2^2} \right) \right] \epsilon^2 v_1^4  \nn \\
	&&	- \left( 1 + \frac{3 \lambda_2 - \lambda_3}{16 \pi^2} \right) y \epsilon \bar{n} n v_1 \nn \\
	&  &  - m_1^4 \left\{ h_1 + \frac{1}{64 \pi^2} \left( \frac{3}{2} - \log \frac{m_1^2 + 3 \lambda_1 v_1^2}{m_2^2} \right) \right\} \nn \\
	& &	- m_2^4 \left( h_2 + \frac{3}{128 \pi^2} \right) - m_{12}^4 \left( h_{12} + \frac{1}{32 \pi^2} \right) \nn \\
	&  & - 2 m_1^2 m_2^2 \left\{ h_3 + \frac{\epsilon^2}{64 \pi^2} \log \left( \frac{m_1^2 + 3 \lambda_1 v_1^2}{m_2^2} \right) \right\} 
		- \frac{1}{2 m_2^2} \left( \frac{\lambda_3}{2} \epsilon v_1^3 - y \bar{n} n \right)^2.
\label{eq:vefflh}
\eea
We observe that the coefficients of the logarithmic term $\log\frac{m_1^2+3\lambda_1 v_1^2}{m_2^2}$  in both expressions are identical to each other.
This implies that the low energy effective potential of Eq.(\ref{eq:veff}) correctly incorporates 
the loop effect of the light scalar.  Since  one can improve the low energy effective potential  Eq.(\ref{eq:veff}) by RG,
we are able to  obtain the RG improved effective potential defined by,
\bea
V^{\tmop{Improved}}_{\tmop{eff}} & = & V_{\tmop{eff}} - V_{\tmop{eff}}^{\tmop{Low}} + V_{\tmop{eff}}^{\tmop{Low \ RGimproved}} ,
\eea
where we set the renormalization scale  $\mu$ equal to $m_2$ in 
 $ V_{\tmop{eff}} $ and $ V_{\tmop{eff}}^{\tmop{Low}} $. 
On the right-hand side, the first two terms $ V_{\tmop{eff}} - V_{\tmop{eff}}^{\tmop{Low}}$ include the loop effect of the heavy scalar. 
Using Eqs.(\ref{RG1}-\ref{eq:hbar}), the solutions of the RG equations,  the RG improved effective potential 
is obtained,
\bea
	V_{\tmop{eff}}^{\tmop{Low \ RGimproved}} 
  		& = & \frac{1}{2} \frac{m_1^2 (m_2)}{\left( 1 + \frac{9 \lambda_1 (m_2)}{16 \pi^2}
  			\log \left( \frac{m_2^2}{m_1^2 + 3 \lambda_1 v_1^2} \right) \right)^{\frac{1}{3}}} 
  			\left( 1 - \frac{9 \lambda_1 (m_2)}{32 \pi^2} \right)v_1^2 \nn \\
		&  &- \frac{1}{2} \frac{m_{12}^2 (m_2)}{\left( 1 + \frac{9 \lambda_1 (m_2)}{16 \pi^2} 
			\log \left( \frac{m_2^2}{m_1^2 + 3 \lambda_1 v_1^2} \right) \right)^{\frac{1}{3}}} 
			\left( 1 - \frac{3 \lambda_1 (m_2)}{16 \pi^2} \right) \epsilon v_1^2\nn \\
		&  & +\frac{1}{4} \frac{\lambda_1 (m_2)}{1 + \frac{9 \lambda_1 (m_2)}{16 \pi^2} 
			\log \left( \frac{m_2^2}{m_1^2 + 3 \lambda_1 v_1^2} \right)} \left( 1 - \frac{27 \lambda_1 (m_2)}{32 \pi^2} \right) v_1^4 \nn \\
		&  & + \frac{1}{4} \frac{\lambda_3 (m_2)}{1 + \frac{9 \lambda_1 (m_2)}{8 \pi^2} 
			\left( 1 - \frac{\lambda_1 (m_2)}{\lambda_3 (m_2)} \right) \log \left( \frac{m_2^2}{m_1^2 + 3 \lambda_1 v_1^2} \right)} 
			\left( 1 - \frac{9 \lambda_1 (m_2)}{8 \pi^2} \left( 1 - \frac{\lambda_1 (m_2)}{\lambda_3 (m_2)} \right) \right) \epsilon^2 v_1^4
			\nn \\
  &  & - m_1^4 \left\{ \bar{h} \left( \sqrt{m_1^2 + 3 \lambda_1 v_1^2}
  \right) + \frac{3}{128 \pi^2} \right\} + 2 \epsilon m_1^2 m_{12}^2 \left\{
  \bar{h} \left( \sqrt{m_1^2 + 3 \lambda_1 v_1^2} \right) + \frac{1}{64 \pi^2}
  \right\} \nn \\
  &  & - \epsilon^2 m_{12}^4 \left\{ \bar{h} \left( \sqrt{m_1^2 + 3 \lambda_1
  v_1^2} \right) \right\} - \epsilon y \bar{n} n v_1 - \frac{1}{2 m_2^2}
  \left( \frac{\lambda_3 \epsilon}{2} v_1^3 - y \bar{n} n \right)^2.
\label{eq:vefflowRGI}
\eea
where the leading logarithmic corrections are summed. $\bar{h}$ denotes the running cosmological constant defined in Eq.(\ref{eq:hbar}) in the
appendix C.
By adding Eq.(\ref{eq:vefflowRGI}) to $ V_{\tmop{eff}} - V_{\tmop{eff}}^{\tmop{Low}} $, one finally obtains the renormalization group improved effective potential,
\bea
V^{\tmop{Improved}}_{\tmop{eff}} & = & 
		\frac{m_1^2 (m_2)}{2} \left[ - \frac{\lambda_3}{32 \pi^2}
		\frac{m_2^2}{m_1^2} + \frac{1 - \frac{9 \lambda_1 (m_2)}{32 \pi^2} }{\left( 1 + \frac{9 \lambda_1 (m_2)}{16 \pi^2}
		\log \left( \frac{m_2^2}{m_1^2 + 3 \lambda_1 v_1^2} \right) \right)^{\frac{1}{3}}} 
		\right] v_1^2 \nn \\
	&  & - \frac{m_{12}^2 (m_2)}{2} \left[ - \frac{5 \lambda_3}{32 \pi^2} + \frac{3 \lambda_2}{16 \pi^2} + \frac{3 \lambda_1}{16 \pi^2} 
		+\frac{1 - \frac{3 \lambda_1 (m_2)}{16 \pi^2} }{\left( 1 + \frac{9 \lambda_1 (m_2)}{16 \pi^2} 
		\log \left(\frac{m_2^2}{m_1^2 + 3 \lambda_1 v_1^2} \right) \right)^{\frac{1}{3}}} \right] \epsilon v_1^2\nn \\
	&  & + \frac{\lambda_1 (m_2)}{4} \left[ \frac{ 1 - \frac{27 \lambda_1 (m_2)}{32 \pi^2}}{1 + \frac{9 \lambda_1 (m_2)}{16 \pi^2} 
		\log \left( \frac{m_2^2}{m_1^2 + 3 \lambda_1 v_1^2}\right)}  \right] v_1^4\nn \\
	&  & + \frac{\lambda_3 (m_2)}{4}\left[ \frac{6 \lambda_2}{16 \pi^2} - \frac{25 \lambda_3}{64 \pi^2} + \frac{9 \lambda_1}{8 \pi^2} 
		\left( 1 - \frac{\lambda_1}{\lambda_3} \right) + \frac{1 - \frac{9 \lambda_1 (m_2)}{8 \pi^2} 
		\left( 1 - \frac{\lambda_1 (m_2)}{\lambda_3 (m_2)} \right) }{1 + \frac{9 \lambda_1 (m_2)}{8 \pi^2} 
		\left( 1 - \frac{\lambda_1 (m_2)}{\lambda_3 (m_2)} \right) \log \left( \frac{m_2^2}{m_1^2 + 3 \lambda_1 v_1^2} \right)} 
		 \right]  \epsilon^2 v_1^4
		\nn \\
	&  & - \left( 1 + \frac{3 \lambda_2 - \lambda_3}{16 \pi^2} \right) y \epsilon \bar{n} n v_1 
		- m_1^4 \left\{ h_1 (m_2) + \frac{1}{64 \pi^2} \left( \frac{3}{2} - \log \frac{m_1^2 + 3 \lambda_1 v_1^2}{m_2^2} \right) \right\}\nn \\
	&  & - m_2^4 \left( h_2 (m_2) + \frac{3}{128 \pi^2} \right) - m_{12}^4 \left( h_{12} (m_2) + \frac{1}{32 \pi^2} \right) \nn \\
	&  & - 2 m_1^2 m_2^2 \left\{ h_3 (m_2) +\frac{\epsilon^2}{64 \pi^2} \log \left( \frac{m_1^2 + 3 \lambda_1 v_1^2}{m_2^2} \right) \right\}
		- \frac{1}{2 m_2^2} \left( \frac{\lambda_3}{2} \epsilon v_1^3 - y \bar{n} n \right)^2.
\label{eq:vimeff}
\eea
In Eq.(\ref{eq:vimeff}), we find the large logarithmic correction to the cosmological constant terms which are  proportional to $m_1^4$ and $m_1^2 m_2^2$ .
The logarithmic correction to the term $m_1^4$ can be interpreted as
the running of the coefficient for  the cosmological constant of
the low energy effective theory (see Eq.(\ref{cosmohbar}) and Eq.(\ref{eq:hbar})).
\begin{eqnarray}
&&  m_1^4 \left\{ h_1 (m_2) + \frac{1}{64 \pi^2} \left( \frac{3}{2} - \log
  \frac{m_1^2 + 3 \lambda_1 v_1^2}{m_2^2} \right) \right\} \nn \\
& = & m_1^4 \left\{h_1 (m_2)+ \frac{3}{128 \pi^2}   - \left( \bar{h} (m_2) -
  \bar{h} \left(\sqrt{m_1^2 + 3 \lambda_1 v_1^2}\right) \right) \right\},
\eea
where,
\bea
\frac{1}{64 \pi^2 } \log \frac{m_1^2 + 3 \lambda_1 v_1^2}{m_2^2} 
    =\bar{h} (m_2) - \bar{h} \left( \sqrt{m_1^2 + 3 \lambda_1 v_1^2} \right) .
\label{cosmohbar}
\end{eqnarray}
The running of  another cosmological constant term which is proportional to $m_1^2 m_2^2 $ in Eq.(\ref{eq:vimeff}) can be also related to that of the  cosmological constant 
term in the low energy effective potential  in Eq.(\ref{eq:veff}).
The corresponding term in Eq.(\ref{eq:vimeff}) is,  
\bea
   &&-2 m_1^2 m_2^2 \left\{ h_3 (m_2) +
 \frac{\epsilon^2}{64 \pi^2} \log \left( \frac{m_1^2 + 3
  \lambda_1 v_1^2}{m_2^2} \right) \right\} \nn \\
 & =& 2 m_1^2 m_{12}^2 \epsilon \left\{- \frac{ h_3 (m_2)}{\epsilon^2}
 -\left( \bar{h} (m_2) - \bar{h} \left( \sqrt{m_1^2 + 3
  \lambda_1 v_1^2} \right) \right) \right\}.
\label{eq:cosmo2}
\eea
In the second line of Eq.(\ref{eq:cosmo2}), we factor out the coefficient $\epsilon m_1^2 m_{12}^2$ so that the comparison to the corresponding term in
 the low energy effective potential in Eq.(\ref{eq:veff}) is feasible. 
\bea
  &&2 \epsilon m_1^2 m_{12}^2 \left\{ \bar{h}(m_2)+ \frac{1}{64 \pi^2} \left( 1 -
  \log \left( \frac{m_1^2 + 3 \lambda_1 v_1^2}{m_2^2} \right) \right) \right\} \nn \\
  &=& 2 \epsilon m_1^2 m_{12}^2 \left\{ \frac{1}{64 \pi^2} + 
\bar{h}(m_2) -\left( \bar{h} (m_2) - \bar{h} \left( \sqrt{m_1^2 + 3 \lambda_1 v_1^2}
  \right)\right) \right\}. 
\label{eq:cosmo3}
\eea
The running of the coefficient for the cosmological constant $\bar{h} (m_2) - \bar{h} \left( \sqrt{m_1^2 + 3 \lambda_1 v_1^2}
  \right) $ in Eq.(\ref{eq:cosmo2}) and Eq.(\ref{eq:cosmo3}) is the same.
Using Eqs.(\ref{cosmohbar}-\ref{eq:cosmo3}), we rewrite the renormalization improved effective potential, Eq.(\ref{eq:vimeff}) as,
\bea
&& V^{\tmop{Improved}}_{\tmop{eff}} \nn \\ 
	&&	=\frac{m_1^2 (m_2)}{2} \left[ - \frac{\lambda_3}{32 \pi^2}
		\frac{m_2^2}{m_1^2} + \frac{1 - \frac{9 \lambda_1 (m_2)}{32 \pi^2} }{\left( 1 + \frac{9 \lambda_1 (m_2)}{16 \pi^2}
		\log \left( \frac{m_2^2}{m_1^2 + 3 \lambda_1 v_1^2} \right) \right)^{\frac{1}{3}}} 
		\right] v_1^2 \nn \\
	&  & - \frac{m_{12}^2 (m_2)}{2} \left[ - \frac{5 \lambda_3}{32 \pi^2} + \frac{3 \lambda_2}{16 \pi^2} + \frac{3 \lambda_1}{16 \pi^2} 
		+\frac{1 - \frac{3 \lambda_1 (m_2)}{16 \pi^2} }{\left( 1 + \frac{9 \lambda_1 (m_2)}{16 \pi^2} 
		\log \left(\frac{m_2^2}{m_1^2 + 3 \lambda_1 v_1^2} \right) \right)^{\frac{1}{3}}} \right] \epsilon v_1^2\nn \\
	&  & + \frac{\lambda_1 (m_2)}{4} \left[ \frac{ 1 - \frac{27 \lambda_1 (m_2)}{32 \pi^2}}{1 + \frac{9 \lambda_1 (m_2)}{16 \pi^2} 
		\log \left( \frac{m_2^2}{m_1^2 + 3 \lambda_1 v_1^2}\right)}  \right] v_1^4\nn \\
	&  & + \frac{\lambda_3 (m_2)}{4}\left[ \frac{6 \lambda_2}{16 \pi^2} - \frac{25 \lambda_3}{64 \pi^2} + \frac{9 \lambda_1}{8 \pi^2} 
		\left( 1 - \frac{\lambda_1}{\lambda_3} \right) + \frac{1 - \frac{9 \lambda_1 (m_2)}{8 \pi^2} 
		\left( 1 - \frac{\lambda_1 (m_2)}{\lambda_3 (m_2)} \right) }{1 + \frac{9 \lambda_1 (m_2)}{8 \pi^2} 
		\left( 1 - \frac{\lambda_1 (m_2)}{\lambda_3 (m_2)} \right) \log \left( \frac{m_2^2}{m_1^2 + 3 \lambda_1 v_1^2} \right)} 
		 \right]  \epsilon^2 v_1^4
		\nn \\
	&  & - \left( 1 + \frac{3 \lambda_2 - \lambda_3}{16 \pi^2} \right) y \epsilon \bar{n} n v_1 
		- m_1^4 \left\{ h_1 (m_2) + \frac{3}{128 \pi^2} - \bar{h} (m_2) + \bar{h} \left( \sqrt{m_1^2 + 3 \lambda_1 v_1^2} \right)
		\right\}\nn \\
	&  & - m_2^4 \left( h_2 (m_2) + \frac{3}{128 \pi^2} \right) - m_{12}^4 \left( h_{12} (m_2) + \frac{1}{32 \pi^2} \right) \nn \\
	&  &- 2 m_1^2 m_{12}^2 \epsilon \left\{ \frac{1}{\epsilon^2} h_3 (m_2) + \bar{h} (m_2) 
		- \bar{h} \left( \sqrt{m_1^2 + 3 \lambda_1 v_1^2} \right) \right\}
		- \frac{1}{2 m_2^2} \left( \frac{\lambda_3}{2} \epsilon v_1^3 - y \bar{n} n \right)^2.
\eea
This completes the derivation of the RG improved effective potential.

\begin{figure}
\begin{center}
\includegraphics[width=8.0cm]{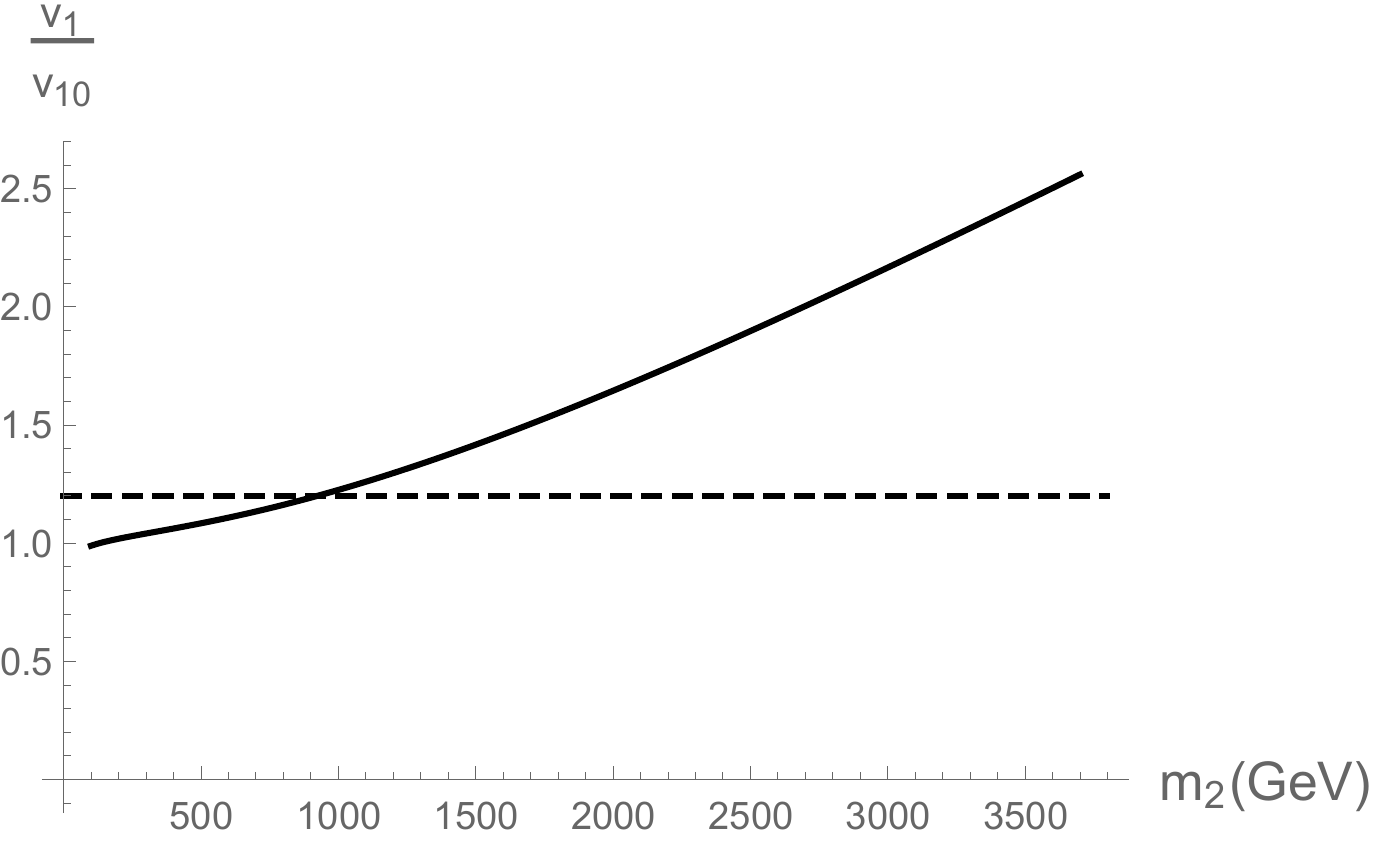}
\end{center}
\caption{The vevs' ratio in Eq.(\ref{eq:vev}) as a function of heavy scalar mass $m_2$. The dashed line corresponds to $\frac{v_1}{v_{10}}=1.2$.}
\label{fig:fig4}
\end{figure}

To illustrate how the vev  depends on the heavy scalar mass, we study the
stationary condition of the effective potential,
\bea
\frac{\partial  V^{\tmop{Improved}}_{\tmop{eff}}}{\partial v_1^2}=0.
\eea
The solution satisfies, 
\bea
&&\frac{v_1}{v_{10}}=
\sqrt{ \Biggl{[}
-\frac{\lambda_3 m_2^2}{32 \pi^2 m_1^2}
+
\frac{1}{\Bigl\{1-
\frac{9 \lambda_1}{16\pi^2} \log \left(\frac{m_1^2+3\lambda_1 v_1^2}{m_2^2}\right) \Bigr\}^{\frac{1}{3}} } \Biggr{]} \Biggl{[}1-
\frac{9 \lambda_1}{16\pi^2} \log \left(\frac{m_1^2+3\lambda_1 v_1^2}{m_2^2}\right) \Biggr{]} } ,\nn \\
&&  v_{10}=\sqrt{-\frac{m_1^2}{\lambda_1}},
\label{eq:vev}
\eea
where we keep the leading logarithmic correction and the correction
proportional to the heavy scalar mass squared. The other small corrections such 
as those suppressed by powers of $\epsilon$ are ignored. 
To study the dependence on the heavy scalar mass of the vev, 
we solve the vev with radiative correction in Eq.(\ref{eq:vev}).  In Fig.(\ref{fig:fig4}), the ratio in Eq.(\ref{eq:vev}) 
is plotted as a function of the heavy scalar mass $m_2$. We fix the parameters as $m_1^2=-(100)^2$(GeV)$^2$ and $\lambda_1=\lambda_3 =1$.  This corresponds to
$v_{10}=100$ (GeV).
The correction to the vev increases as the heavy scalar mass increases. 
If we require the  correction is within  $20\%$  compared to the vev without the radiative correction, the upper bound on the heavy scalar mass 
is about $1000$ (GeV).  
\section{Summary and Discussion}
In this paper, we have studied a model consists of a light scalar and a heavy scalar. In the model, we introduce the Yukawa interaction for neutrinos coupled to the heavy scalar.
We also add the cosmological constant terms which are related to the mass parameters of the model. To derive the low energy effective potential, we integrate 
 both light and heavy scalars out with the constant expectation value of the light scalar. This is realized by introducing generating functional with the source only for the light scalar. Then performing the Legendre transformation, the effective potential of the light scalar is obtained. In this way, one particle irreducibility for the light scalar is maintained and the effects of the diagrams where the heavy scalar is exchanged can be also included. 

We have found that the effective potential is  independent of the renormalization scale approximately. 
 The large logarithmic correction for the heavy scalar loop is suppressed by setting the renormalization scale equal to the heavy scalar mass. With this choice of the renormalization
scale,  the large logarithmic corrections in  the effective potential originate from the loop effect of the light scalar. We derive the low energy effective potential with the tree level matching and one-loop calculation of the light scalar. As a result, we find that the logarithmic corrections in the both effective potential are the same. As for the low energy effective potential, we perform the RG improvement  by  summing  the leading logarithmic corrections. 
Since the loop correction due to the heavy scalar is absent in the low energy effective potential, 
we add this correction as the difference of the two effective potentials. 

We found that the effective Yukawa  coupling  $y_{\rm{eff}}$ between 
 the  light scalar and the Dirac neutrino is inversely proportional to the heavy scalar mass squared as
$y_{\rm{eff}}\simeq y \frac{m_{12}^{2}}{m_{2}^{2}}$ and is naturally suppressed. As for the dimension six operators, 
the six point interaction for the light scalars and the four Fermi  interaction for the neutrinos are also generated.
A part of the radiative corrections to the light scalar mass squared is proportional to the heavy scalar mass squared. 
We numerically study the effect to the vev using the stationary condition of the potential.
The vev depends on the heavy scalar mass and one can set the upper bound on the mass
by demanding the radiative correction to the vev should be limited within a certain range. 

The cosmological constants also suffer from the large contribution
proportional to the fourth power of the heavy scalar mass.   
As for the renormalizable coupling such as the quartic interaction of the light scalar,  
the contribution suppressed by a factor of $\epsilon^2$ appear. This implies the renormalizable coupling constant of the low energy effective potential is also sensitive to the coupling and the mass of the full theory.  
  
For the future work, we will extend the present analysis to the realistic model based upon  the standard model  gauge group  such as Davidson and Logan Model \cite{Davidson:2009ha}, and derive the low energy effective theory \cite{Jenkins:2017dyc, Jenkins:2017jig}. With the realistic model, one can constrain the parameters of the full theory such
as the heavy scalar mass and the small mass mixing term for two scalars from the correction to the vev of the light scalar and the effective Yukawa coupling to neutrinos. 
One can also obtain the constrains on the coefficients of the cosmological constants.
\\
\vspace{1cm}
\\
\noindent
{\bf Acknowledgement}\\
The work of T.M. is supported by Japan Society for the Promotion of Science (JSPS) KAKENHI Grant Number JP17K05418 .
\appendix
\section{The counter terms for the full theory}
In this appendix, we  determine the counter terms for the full theory. We also derive the RG equation and examine the renormalization point 
independence of the effective couplings and masses.
Since there is no wave function renormalization from one-loop contribution of scalar fields, it is 
sufficient to know the counter terms for the effective potential. 
The Lagrangian density for the full theory is;
\bea
	\mathcal{L} & = & - \frac{1}{2}  \left( \sum_{i = 1}^2 Z_i \rho_i \Box \rho_i + \sum_{i, j = 1}^2 \rho_i^2 Z_{\tmop{mij}} m_j^2 
		+ \frac{\mu^{2 \eta}}{2} \sum_{I = 1}^3 \left( \sum_{i = 1}^2 \rho_i^4 Z_{\lambda \tmop{iI}} \lambda_I 
		+ \rho_1^2 \rho_2^2 Z_{\lambda_{} 3 I} \lambda_I \right) \right) \nn \\
		&- &  (Z_y y \mu^{\eta}  \bar{n} n + Z_{12} m_{12}^2 \rho_1) \rho_2 + \bar{n} i \not{\partial} n Z_n 
		+\mu^{d - 4} Z_{h_1} h_1 m_1^4 + \mu^{d - 4} Z_{h_2} h_2 m_2^4 + \mu^{d - 4} Z_{h_{12}} h_{12} m_{12}^4 \nn \\
	&=& - \frac{1}{2}  \left( \sum_{i = 1}^2 \rho_{0i} \Box \rho_{0i} + \sum_{i  = 1}^2 \rho_{0i}^2 m_{0i}^2 
		+ \frac{1}{2} \left( \sum_{i = 1}^2 \rho_{0i}^4 \lambda_i + \rho_{01}^2 \rho_{02}^2 \lambda_{03} \right) \right) \nn \\
		& - &  ( y_0  \bar{n}_0 n_0 + m_{012}^2 \rho_{01} ) \rho_{02} 
		+\mu^{d - 4} Z_{h_1} h_1 m_1^4 + \mu^{d - 4} Z_{h_2} h_2 m_2^4 + \mu^{d - 4} Z_{h_{12}} h_{12} m_{12}^4.
\eea
The counter terms are given by,
\begin{eqnarray*}
  \mathcal{L}_c &=&  - \frac{1}{2}  \sum_{i = 1}^2 (Z_i - 1) \rho_i \Box \rho_i 
  -\frac{1}{2} \sum_{i, j = 1}^2 \rho_i^2 (Z_{\tmop{mij}} - \delta_{\tmop{ij}}) m_j^2 - ((Z_y - 1)_{} y \mu^{\eta}  \bar{n} n + (Z_{12} - 1) m_{12}^2
  \rho_1) \rho_2 \nn \\
  &-&  \frac{\mu^{2 \eta}}{4} \sum_{I = 1}^3 \left( \sum_{i = 1}^2 \rho_i^4
  (Z_{\lambda_{} \tmop{iI}} - \delta_{\tmop{iI}}) \lambda_I + \rho_1^2
  \rho_2^2  (Z_{\lambda_{} 3 I} - \delta_{3 I}) \lambda_I \right) \nn \\
  & + & i \bar{n} \not{\partial} n (Z_n - 1) - \mu^{d - 4} (Z_{h_1} - 1) h_1
  m_1^4 - \mu^{d - 4} (Z_{h_2} - 1) h_2 m_2^4 - \mu^{d - 4} (Z_{h_{12}} - 1)
  h_{12} m_{12}^4. 
\end{eqnarray*}
Next we compute the one-loop effective potential.
\bea
 V_{\tmop{eff}}^{1 \tmop{loop}} & = & - i \frac{1}{2 V^{d - 1} T} \log \det \left( \left.  \frac{- \delta^2 S_{\tmop{tree}}}{\delta \rho_i (x) \delta \rho_j (y)}
			\right|_{\rho_i = v_i \mu^{- \eta}} \right) + V_c \nn \\
&= &  \frac{1}{2} \int \frac{d^d k}{(2\pi)^d i}  \log \Bigl[ \left( m_1^2 +  \left( 3 \lambda_1 v^2_1 + \frac{\lambda_3}{2} v_2^2 \right)-k^2 \right)
 \left( m_2^2 +  \left( 3 \lambda_2 v_2^2 + \frac{\lambda_3}{2} v_1^2 \right) - k^2 \right)  \nn \\
 &-& (m_{12}^2+  v_1 v_2 \lambda_3 )^2 \Bigr]+ V_c. 
\eea
where $V_c$ is the countertems for effective potential given by,
\bea
V_c & = & \frac{\mu^{-2\eta}}{2} \Bigl[ \{ (Z_{m 11} - 1) m_1^2 + Z_{m 12} m_2^2 \} v_1^2 
		+ \{ (Z_{m 22} - 1) m_2^2 + Z_{m 21} m_1^2 \} v_2^2 + 2 (Z_{12} - 1) m_{12}^2  v_1 v_2 \Bigr]  \nn \\
		& + &  \mu^{-2 \eta} \frac{(Z_{\lambda 11} - 1) \lambda_1 + Z_{\lambda12} \lambda_2 + Z_{\lambda 13} \lambda_3}{4} v_1^4 
		+ \mu^{-2 \eta} \frac{Z_{\lambda 21} \lambda_1 + (Z_{\lambda 22} - 1) \lambda_2 
		+ Z_{\lambda 23} \lambda_3}{4}v_2^4 \nn \\
		& + &   \mu^{-2 \eta} \frac{Z_{\lambda 31} \lambda_1 
		+ Z_{\lambda 32} \lambda_2 + (Z_{\lambda 33} - 1) \lambda_3}{4} v_1^2 v_2^2+ (Z_y - 1)y   \bar{n} n v_2 \nn \\
		&- &  \mu^{-2 \eta} \{ (Z_{h_1} - 1) h_1 m_1^4 + (Z_{h_2} - 1) h_2 m_2^4 +  (Z_{h_{12}} - 1) h_{12} m_{12}^4 \}.
\eea
The divergent parts can be extracted by expanding the logarithmic terms up to $(m_{12}^2+ v_1 v_2 \lambda_3 )^2$.
\bea
	V_{\tmop{eff}}^{1 \tmop{loop}} 
		& \simeq & \frac{1}{2} \int \frac{d^d k}{(2 \pi)^d i}  \left\{ \log ( m_1^2 
			+3 \lambda_1 v^2_1 + \frac{\lambda_3}{2} v_2^2  - k^2) 
			+ \log ( m_2^2 +  
			3 \lambda_2 v_2^2 + \frac{\lambda_3}{2} v_1^2  - k^2 ) \right\} \nn \\
			& - & \frac{1}{2} \int \frac{d^d k}{(2 \pi)^d i} 
			\frac{ (m_{12}^2+  v_1 v_2 \lambda_3 )^2}{(m_1^2 +  3 \lambda_1 v^2_1 + \frac{\lambda_3}{2} v_2^2  - k^2)(m_2^2 +  3 \lambda_2 v_2^2 + \frac{\lambda_3}{2} v_1^2  - k^2)} +V_c
 +...\nn \\
\eea
We collect only the one loop divergences and the counter terms as follows.
\bea
	&& V_{\tmop{eff}}^{1 \tmop{loop} \tmop{div} }+V_c \nn \\
		& = & - \frac{C_{\tmop{UV}}}{64 \pi^2} \{ m_1^4 + m_2^4 + 2 m_{12}^4 \} 
			- \frac{C_{\tmop{UV}}}{32 \pi^2}  \left( 3 m_1^2 \lambda_1 + m_2^2 \frac{\lambda_3}{2} \right) v_1^2 \nn \\
			&-& \frac{C_{\tmop{UV}}}{32 \pi^2}  \left( 3 m_2^2 \lambda_2 + m_1^2 \frac{\lambda_3}{2} \right) v_2^2 
			- \frac{C_{\tmop{UV}}}{16 \pi^2} m_{12}^2 \lambda_3 v_1 v_2 \nn \\
			& - & \frac{C_{\tmop{UV}}}{64 \pi^2} \left\{ \left( 9 \lambda_1^2 + \frac{\lambda_3^2}{4} \right) v_1^4 
			+ \left( 9 \lambda_2^2 + \frac{\lambda_3^2}{4} \right) v_2^4 + 3 (\lambda_1 + \lambda_2) \lambda_3 v^2_1 v_2^2 
			+ 2 \lambda_3^2 v_1^2 v_2^2 \right\} \nn \\ 
& + & \frac{\mu^{-2\eta}}{2} \Bigl[ \{ (Z_{m 11} - 1) m_1^2 + Z_{m 12} m_2^2 \} v_1^2 
		+ \{ (Z_{m 22} - 1) m_2^2 + Z_{m 21} m_1^2 \} v_2^2 + 2 (Z_{12} - 1) m_{12}^2  v_1 v_2 \Bigr]  \nn \\
		& + &  \mu^{-2 \eta} \frac{(Z_{\lambda 11} - 1) \lambda_1 + Z_{\lambda12} \lambda_2 + Z_{\lambda 13} \lambda_3}{4} v_1^4 
		+ \mu^{-2 \eta} \frac{Z_{\lambda 21} \lambda_1 + (Z_{\lambda 22} - 1) \lambda_2 
		+ Z_{\lambda 23} \lambda_3}{4}v_2^4 \nn \\
		& + &   \mu^{-2 \eta} \frac{Z_{\lambda 31} \lambda_1 
		+ Z_{\lambda 32} \lambda_2 + (Z_{\lambda 33} - 1) \lambda_3}{4} v_1^2 v_2^2+ (Z_y - 1)y   \bar{n} n v_2 \nn \\
		& - &  \mu^{-2 \eta} \{ (Z_{h_1} - 1) h_1 m_1^4 + (Z_{h_2} - 1) h_2 m_2^4 +  (Z_{h_{12}} - 1) h_{12} m_{12}^4 \}.
\eea
We require that the counter terms subtract all the divergences,
\bea
V_c & = & - V_{\tmop{eff}}^{1 \tmop{loop} \tmop{div}}.
\eea
Then one can determine the Z factors of the counter terms as follows,
\bea
	Z_{m \tmop{ii}} &=& 1 + \frac{3 C_{\tmop{UV}}}{16 \pi^2} \lambda_i, \quad
	Z_{m 12} = Z_{m 21} = \frac{C_{\tmop{UV}}}{32 \pi^2} \lambda_3,  \label{eq:Zm21}\\
	Z_{12} &=& 1 + \frac{C_{\tmop{UV}}}{16 \pi^2} \lambda_3, 	\quad Z_{\lambda \tmop{ii}}= 1 + \frac{C_{\tmop{UV}}}{16 \pi^2} 9 \lambda_i,  \\
	Z_{\lambda i 3} &=& \frac{C_{\tmop{UV}}}{64 \pi^2} \lambda_3,  \quad
		Z_{\lambda 3 i} =  \frac{C_{\tmop{UV}}}{16 \pi^2} 3\lambda_3, \quad 
	Z_{\lambda 33}  = 1+ \frac{C_{\tmop{UV}}}{8 \pi^2} \lambda_3,  \\
     Z_y  &=&  1, \quad Z_{\lambda 12} =Z_{\lambda 21} = 0, \\
(Z_{h_i} - 1) h_i&=&- \frac{C_{\tmop{UV}}}{64 \pi^2}, \quad  (Z_{h_{12}} - 1) h_{12} =  - \frac{C_{\tmop{UV}}}{32 \pi^2},
\label{eq:fullcounter}
\eea
where $i=1,2$.  This completes the derivation of the Z factors in the counter terms for the full theory.
The RG equations for coupling constants can be also derived as,
\bea
  \mu \frac{d \lambda_3}{d \mu} & = &
 \frac{1}{8 \pi^2} (2 \lambda_3 + 3 \lambda_2 + 3 \lambda_1) \lambda_3, \label{eq:rg1}\\
  \mu \frac{d \lambda_i}{d \mu} & = & \frac{1}{8 \pi^2} 9 \lambda_i^2 +
  \frac{1}{32 \pi^2} \lambda_3^2,  \quad (i = 1, 2).
\eea
The coefficients of the cosmological constants and the mass parameters satisfy the following RG 
equations.
\begin{eqnarray}
	\mu \frac{d h_1}{d \mu} & = & \frac{- 1}{16 \pi^2} \left( \frac{1}{2} + 12 \lambda_1 h_1 + 2 \lambda_3 h_3 \right), \\
	\mu \frac{d h_2}{d \mu} & = & \frac{- 1}{16 \pi^2} \left( \frac{1}{2} + 12 \lambda_2 h_2 + 2 \lambda_3 h_3 \right), \\
	\mu \frac{d h_3}{d \mu} & = & \frac{- 1}{16 \pi^2} (\lambda_3 (h_1 + h_2) + 6 (\lambda_2 + \lambda_1) h_3), \\
	\mu \frac{d h_{12}}{d \mu} & = & - \frac{1}{16 \pi^2} (1 + 4 \lambda_3 h_{12}), \\
	\mu \frac{d m_1^2}{d \mu} & = &  \frac{3}{8 \pi^2} \lambda_1 m_1^2 + \frac{1}{16 \pi^2} \lambda_3 m_2^2, \\
	\mu \frac{d m_2^2}{d \mu} & = &  \frac{3}{8 \pi^2} \lambda_2 m_2^2 + \frac{1}{16 \pi^2} \lambda_3 m_1^2, \\
	\mu \frac{d m_{12}^2}{d \mu} & = & \frac{\lambda_3}{8 \pi^2} m_{12}^2.
	\label{eq:rg18}
\end{eqnarray}
Using the RG equations from Eq.(\ref{eq:rg1}) to Eq.(\ref{eq:rg18}), one can
study the renormalization point independence of the effective masses and the effective 
coupling constants in the one-loop effective potential of  Eq.(\ref{Veffective}).
\bea
	\mu \frac{d m_{1 \tmop{eff}}^2}{d \mu} & = & m_1^2\frac{3\lambda_1}{8 \pi^2}  + m_2^2 \frac{\lambda_3  }{16 \pi^2} 
		- m_1^2  \frac{3 \lambda_1 }{8 \pi^2} - \frac{\lambda_3 m_2^2}{16 \pi^2} = 0, \\
	\mu \frac{d (m_{12 \tmop{eff}}^2 \epsilon)}{d \mu} & = & \frac{m_{12}^4}{m_2^2} \left\{ \frac{\lambda_3}{4 \pi^2} 
		- \frac{3\lambda_2 }{8 \pi^2}  + \frac{3\lambda_2 }{8 \pi^2}  - \frac{\lambda_3 }{4 \pi^2} 
		+ \frac{\lambda_3}{16 \pi^2}  \frac{m_1^2}{m_2^2} \right\}
		=\frac{\lambda_3 m_{12}^2 \epsilon}{16 \pi^2} \left( \frac{m_1^2}{m_2^2} \right), \\
	\mu \frac{d \lambda_{1 \tmop{eff}}}{d \mu} & = & \frac{9 \lambda_1^2}{8 \pi^2} + \frac{\lambda_3^2}{32 \pi^2}  
		+ \lambda_1 \left\{ - \frac{9 \lambda_1 }{8 \pi^2} - \frac{1}{32 \pi^2} \frac{\lambda_3^2 }{\lambda_1} \right\} = 0, \\
	\mu \frac{d (\lambda_{3 \tmop{eff}} \epsilon^2)}{d \mu} & = & \epsilon^2 
		\left\{ \frac{\lambda_3 (2 \lambda_3 + 3 \lambda_2 + 3 \lambda_1 + 3 \lambda_2 - 4 \lambda_3 - 3 \lambda_1 
		+ 2 \lambda_3 - 6 \lambda_2)}{8 \pi^2}\right\} - \frac{\lambda_3}{8 \pi^2} \frac{m_1^2}{m_2^2}  \nn \\
		& = & - \frac{\lambda_3 \epsilon^2}{8 \pi^2} \left( \frac{m_1^2}{m_2^2} \right),\\
		\mu \frac{d (y_{\tmop{eff}} \epsilon)}{d \mu} & = & \epsilon y 
		\left\{ \frac{3 \lambda_2 - \lambda_3 + \lambda_3 - 3 \lambda_2}{8 \pi^2} \right\} 
		- \frac{\epsilon y \lambda_3 }{16 \pi^2} \frac{m_1^2}{m_2^2} 
		= - \frac{\lambda_3 y \epsilon}{16 \pi^2} \left( \frac{m_1^2}{m_2^2} \right).
\eea
One finds all of the effective masses and the couplings are renormalization point
independent by ignoring the sub-leading corrections $O(\frac{m_1^2}{m_2^2})$. The remaining dependence is due to the
truncation of these suppressed contribution in the derivation of  the effective potential. 
The cosmological constants of Eq.$(\ref{Vcosmo})$ is written as follows,
\begin{eqnarray}
  V_{\tmop{cosmo}} & = & - \frac{m_{12}^4}{32 \pi^2}  \left( 1 - \log
  \frac{m_2^2}{\mu^2} \right) - \frac{m_1^4}{64 \pi^2}  \left( \frac{3}{2} -
  \log \left( \frac{m_1^2 + 3 \lambda_1  v_1^2 }{\mu^2}
  \right) \right) \nn \\
  &  & - \frac{m_2^4}{64 \pi^2}  \left( \frac{3}{2} - \log
  \frac{m_2^2}{\mu^2} \right)  - h_1 m_1^4 - h_2 m_2^4 - h_{12} m_{12}^4 - 2 h_3 m_1^2 m_2^2.
\end{eqnarray}
The renormalization point independence of $ V_{\tmop{cosmo}}$ is explicitly shown below. 
\begin{eqnarray}
  \mu \frac{d V_{\tmop{cosmo}}}{d \mu} & = & \frac{2 \lambda_3 h_3}{16 \pi^2}
  (m_1^4 + m_2^4) - 2 (h_1 + h_2) m_1^2 \frac{1}{16 \pi^2} \lambda_3 m_2^2\nn \\
  &  & + \frac{2}{16 \pi^2} (\lambda_3 (h_1 + h_2)) m_1^2 m_2^2 - h_3
  \frac{2}{16 \pi^2} \lambda_3 (m_2^4 + m_1^4) \nn \\
  & = & 0.
\end{eqnarray}
\section{Derivation of $V_{\rm{eff}}^{\rm{Low}}$ in Eq.(\ref{eq:veff})}
In this appendix, we give an outline of the derivation of Eq.(\ref{eq:veff}).
In Eq.(\ref{eq:veff}), 
$V_{\tmop{eff}}^{1 \tmop{loop}}$ is given by,
\bea
V_{\tmop{eff}}^{1 \tmop{loop}}
		& = & \frac{1}{2} \int \frac{d^d k}{(2 \pi)^d i} \log (m_1^{\prime 2} + 3 \lambda_1^\prime v_1^2- k^2) \nn \\
&=& -\frac{(m_1^{\prime 2} + 3 \lambda_1^\prime v_1^2)^2}{64 \pi^2}\left(C_{UV}+\frac{3}{2}-
\log(m_1^{\prime 2} + 3 \lambda_1^\prime v_1^2) \right),
\label{eq:Veff1loop}
\eea
where the divergence is denoted by 
$
C_{UV}=\frac{1}{\eta}-\gamma+\log 4\pi
$
and the coupling constant and the mass are defined as, 
\bea
m_1^{\prime 2}=m_1^2-\epsilon m_{12}^2 , \quad \lambda_1^\prime= \lambda_1+\epsilon^2 (\lambda_3-\lambda_1). 
\eea
The divergent part is extracted  from Eq.(\ref{eq:Veff1loop}) and is given as,
 \bea
V_{\tmop{eff}}^{1 \tmop{loop \ div.}}&=& 
-\frac{(m_1^{\prime 2} + 3 \lambda_1^\prime v_1^2)^2}{64 \pi^2} C_{UV} \nn \\
&=&-\frac{(m_1^{2} -\epsilon m_{12}^2)^2+6 (m_1^{2} -\epsilon m_{12}^2) ( \lambda_1 +\epsilon^2 (\lambda_3-\lambda_1) ) v_1^2
+9 ( \lambda_1 +\epsilon^2 (\lambda_3-\lambda_1) )^2 v_1^4}{64 \pi^2} C_{UV}.\nn \\
\eea
Then the divergence can be subtracted by the counter terms,
\bea
  V^c_{\tmop{eff}} & = & \frac{1}{2} (Z_{m'_1} - 1)
 (m_1^2-\epsilon m^2_{12} ) \mu^{- 2 \eta} v_1^{2} + \frac{\lambda_1'}{4} (Z_{\lambda_1'} - 1)
  v_1^{4} \mu^{- 2 \eta} 
-(Z_{\bar{h}}-1) \mu^{-2 \eta } \bar{h}( m_1^2- \epsilon m^2_{12} )^2.\nn \\
\nn \\
\label{eq:countertermforveff}
\eea
where the $Z$ factors are determined so that the counter terms satisfy
$V^c_{\tmop{eff}}=-V_{\tmop{eff}}^{1 \tmop{loop \ div.}}$.
\bea
 &&Z_{m'_1} - 1 =  \frac{3 \lambda'_1 C_{\tmop{UV}}}{16 \pi^2} ,\quad
  Z_{\lambda_1'} - 1 =  \frac{9 C_{\tmop{UV}}}{16 \pi^2} \lambda_1'. \\
&& (Z_{\bar{h}}-1)\bar{h}= -\frac{1}{64\pi^2} C_{UV} .
\label{eq:zfactorloweff}
\eea
By adding the counter terms to the tree and one-loop contribution,
the finite effective potential is obtained,  
\bea
&& V_{\rm{eff}}+V_{\tmop{eff}}^{1 \tmop{loop}}+V^c_{\tmop{eff}} \nn \\
&&=- h_1 {m_1}^4  -h_2 {m_2}^4 - 2 h_3 {m_1}^2{m_2}^2 - {h_{12}}m_{12}^4 - \frac{1}{2 m_2^2} \left( \frac{\epsilon \lambda_3 {v_1}^3}{2} - y \bar{n} n \right)^2  \nn \\
&&+\frac{m_1^2 - \epsilon m_{12}^2}{2} {v_1}^2 + \frac{\lambda_1 + \epsilon^2 \lambda_3 }{4}{v_1}^4 -\epsilon y \bar{n} n {v_1}  \nn \\
&& + \frac{(m_1^{2} -\epsilon m_{12}^2+ 3( \lambda_1 +\epsilon^2 (\lambda_3-\lambda_1)) v_1^2)^2}{64 \pi^2}\left(
\log\frac{m_1^{2} -\epsilon m_{12}^2+ 3( \lambda_1 +\epsilon^2 (\lambda_3-\lambda_1)) v_1^2}{\mu^2} -\frac{3}{2} \right). \nn \\
\label{eq:veff0}
\eea
To obtain the final form of the effective potential Eq.(\ref{eq:veff}) from Eq.(\ref{eq:veff0}),  we keep the terms such as $m_1^2
v_1^2 \epsilon^0$,  $m_{12}^2 v_1^2 \epsilon$ , $ v_1^4$ and $\epsilon^2 v_1^4$. 
 About the cosmological constant terms,
 we keep the terms of the forms $m_1^4 \epsilon^0$, $m_1^2
m^2_{12} \epsilon $ and $m_{12}^4 \epsilon^2 $. 

\section{ RG equation and its solutions  for low energy effective theory}
In this appendix, the RG equations for the coupling ,mass, and cosmological constant of  the low energy effective action in
Eq.(\ref{eqlowL})
are obtained. The solutions in the leading logarithmic approximation are also derived.  
The relations between the renormalized quantities and the bare ones are given as follows;
\bea
\lambda^{\prime }_{10} =  Z_{\lambda^\prime_1} \lambda'_1 \mu^{2 \eta} ;\quad
m_{10}^{\prime 2} =  Z_{m_1^\prime} m_1^{\prime 2} ;\quad
\bar{h}_0 m^{\prime 4}_{10}  =  Z_{\bar{h}} \bar{h} m_1^{\prime 4} \mu^{- 2 \eta}
\eea
where $Z$ factors are given by, 
\bea
  Z_{\lambda_1^\prime} =  1 + \frac{9 C_{\tmop{UV}}}{16 \pi^2} \lambda_1^\prime,
  \quad  Z_{m_1^\prime}  =  1 + \frac{3 \lambda_1^\prime C_{\tmop{UV}}}{16 \pi^2},
 \quad (Z_{\bar{h}} - 1) \bar{h}  =  - \frac{C_{\tmop{UV}}}{64 \pi^2}.
\eea
Using the relations, the RG equations are derived as,
\bea
& &\mu \frac{d m^{\prime 2}_1 (\mu)}{d \mu} - m^{\prime 2}_1 (\mu) \frac{3 \lambda_1^\prime}{8 \pi^2} = 0, \\
& &\mu \frac{d \lambda_1^\prime (\mu)}{d \mu} - \frac{9 \lambda_1^{\prime 2}}{8 \pi^2}  = 0, \\
& &\mu \frac{d \bar{h}}{d \mu} + \frac{1}{32 \pi^2}  =  0.
\eea
By solving the RG equations, we obtain the RG improved couplings and masses. 
\bea
\lambda_1^\prime (\mu_0) & = & \frac{\lambda_1^\prime (\mu)}{1 + \frac{9 \lambda_1^\prime (\mu)}{16 \pi^2} \log \frac{\mu^2}{\mu_0^2}},
 \label{RG1}\\
m_1^{\prime 2} (\mu_0) & = & 
		\frac{m^{\prime 2}_1 (\mu)}{\left( 1 + \frac{9 \lambda_1^\prime (\mu)}{16 \pi^2} \log \frac{\mu^2}{\mu_0^2} \right)^{\frac{1}{3}}}, \\
\bar{h} (\mu_0) & = & \bar{h} (\mu) - \frac{1}{64 \pi^2 } \log \frac{\mu_0^2}{\mu^2} .
\label{eq:hbar}
\eea

\end{document}